\documentclass{osa-article}

\journal{osajournal}

\articletype{Research Article}

\usepackage{enumitem}
\usepackage{multirow}
\usepackage{lineno}

\newcommand{\revision}[1]{\textcolor{red}{#1}}
\renewcommand{\revision}[1]{{#1}} 

\graphicspath{%
  {./ImagesArticleConvert/}%
  {./ImagesAppendixConvert/}%
}

\begin{document}

\title{Near ground horizontal high resolution $C_n^2$ profiling from Shack-Hartmann slope and scintillation data}

\author{C. Sauvage,\authormark{1,2,6}  C. Robert,\authormark{1,5}  
L. M. Mugnier,\authormark{1}  
J.-M. Conan,\authormark{1}  
J.-M. Cohard,\authormark{3}  
K.-L. Nguyen,\authormark{1}  
M. Irvine,\authormark{4} 
and J.-P. Lagouarde\authormark{4}}

\address{%
\authormark{1}DOTA, ONERA, Universit\'e Paris Saclay, 92320 Ch\^atillon, France\\
\authormark{2}LTCI, T\'el\'ecom Paris, Institut Polytechnique de Paris, 91120 Palaiseau, France\\
\authormark{3}UGA (Universit\'e de Grenoble Alpes) CS 40 700, 38058 Grenoble Cedex 9, France\\
\authormark{4}INRAE UMR 1391 Interactions Sol Plante Atmosphère CS 20032, 33882 Villenave d'Ornon Cedex, France
}
\email{\authormark{5}clelia.robert@onera.fr \\ \authormark{6} chloe.sauvage.phd@gmail.com}



\begin{abstract}
CO-SLIDAR is a very promising technique for the metrology of near ground $C_n^2$ profiles.
It exploits both phase and scintillation measurements obtained with a dedicated wavefront sensor and allows profiling on the full line of sight between pupil and sources.
This technique is applied to an associated instrument based on a mid-IR Shack-Hartmann wavefront sensor, coupled to a \revision{0.35\,m} telescope, which observes two cooperative sources. This paper presents the first comprehensive description of the CO-SLIDAR method in the context of near ground \revision{optical turbulence} metrology. 
It includes the presentation of the physics principles underlying the measurements, of our unsupervised $C_n^2$ profile reconstruction strategy together with the error bar estimation on the reconstructed values.
The application to data acquired in a heterogeneous rural landscape during an experimental campaign in Lannemezan (France) demonstrates the ability \revision{to obtain  profiles with a sampling pitch of about 220\,m} over a 2.7\,km line of sight. The retrieved $C_n^2$ profiles are presented and their variability in space and time is discussed.
\end{abstract}

\section{Introduction}
\label{sect:intro}  
Sensible and latent heat fluxes are the main drivers for atmospheric boundary layer dynamics. They are highly variable in time and space \cite{Cheinet2009}, and their spatial variability is particularly difficult to measure under heterogeneous conditions in the field \cite{ Maronga2013}. 
The sensible heat flux can be deduced from the turbulent structure parameter of temperature $C_T^2$, which in turn can be computed from the $C_n^2$, i.e., the refractive index structure parameter that characterizes optical turbulence induced by variations in pressure, temperature and humidity in the atmosphere. The $C_n^2$ profile characterizes the local strength of the turbulence.

Interactions between the climate and the biosphere are governed by a set of processes operating from the local to the global scale. 
The interval ranging from \revision{$hm$ to a few tens of $km$}, called the ``landscape'' scale, is particularly suitable to study continental impacts of heat fluxes on the climate for hydrological~\cite{Poisson2016} as for agricultural~\cite{Lagouarde2000, Lagouarde2002} or urban issues~\cite{lagouarde_monitoring_2006}.
Among other measurement techniques, scintillometry has been recognized, for a number of years, as a suitable technology to characterize the role of the interface between the lower atmosphere and the biosphere at km$^2$ scales~\cite{de_bruin_scintillation_1995,meijninger_determination_2002,meijninger_determination_2002-1,lagouarde_monitoring_2006,Moene2009,guyot_combined_2009,guyot_long-term_2012,ward_multi-season_2013}.
Indeed, scintillometry fills the gap between local station networks which are demanding in terms of maintenance and monitoring, and airborne instrumentation  which is convenient for larger scales, difficult to implement and incompatible with continuous monitoring.
Over the last decades, scintillometry has been a key remote sensing technique to access average turbulent fluxes. But none of these techniques (eddy-covariance stations, scintillometry, aircraft measurements) \revision{allow} to profile the $C_n^2$ and thus the heat fluxes at hm$^2$ scales where homogeneous surface conditions are fulfilled. One therefore lacks an instrument able to profile remotely the turbulence characteristics in the surface layer.

The evaluation of the influence of turbulence on near ground applications such as optical communications, imaging and remote sensing, relies also on the knowledge of $C_n^2$ distribution near the ground. Many experiments have thus been carried out along a horizontal line of sight over different ground surfaces. In these studies \cite{Phillips1981, Thiermann1995, Beaumont1997, Biswas2000,bose-pillai_estimation_2018}, the $C_n^2$ profile is, however, often assumed to be uniform along the line of sight, which is far from the reality. The knowledge of the heterogeneous $C_n^2$ profile along a horizontal path is therefore a key issue for optical instruments \cite{Ingensand2008}.

Furthermore, we recall that $C_n^2$ profiling techniques have been the object of intensive developments for non horizontal applications (ground astronomy, ground-space links, etc.). 
The $C_n^2$ profiling with optical instruments exploits one or several of the following physical properties that can be stated as:
\begin{enumerate}[label=(\roman*)]
    \item optical turbulence observed in a reception plane/pupil transverse to the incoming beam has two types of signature: phase effects and amplitude alias scintillation effects,
    \item phase effects can be approximated as a mere integral of phase perturbations over layers along $z$, where $z$ is the distance to pupil plane along the line of sight,
    \item scintillation effects have a spatial structure that depends on $z$: characteristic size $\sqrt{\lambda z}$ (\revision{see footnote}\footnotemark[1]),
\footnotetext[1]{Expression is given for sources at infinity but can be easily generalized for sources at finite distance.} where $\lambda$ is the wavelength, and a sensitivity that increases with $z$ (layers near the instrument do not contribute to scintillation),
    \item phase and scintillation patterns induced by a given turbulent layer at $z$  are identical for 2 sources separated by an angle $\Theta$ apart from a translation $\Theta z$ (\revision{see footnote}\footnotemark[1]). 
\end{enumerate}

A first strategy for $C_n^2$ profile estimation relies on slope measurements on several apertures. Due to (ii), profiling requires triangulation, that is, the comparison of the patterns from two or more sources based on (iv) (SLODAR \cite{Butterley2006, Wilson2002}, PBL/PML \cite{Ziad2013}, S-DIMM+ \cite{Scharmer2010}). A second strategy called SCIDAR (SCIntillation Detection And Ranging) consists in using triangulation on a double source applied to pupil plane scintillation patterns \cite{Rocca1974}. Due to (iii) SCIDAR is, however, not sensitive to layers close to the pupil, which prompted the development of the generalized SCIDAR, denoted G-SCIDAR~\cite{Avila1997, Fuchs1998}, 
in which detection is performed below the pupil plane, and of its recent variation Stereo-SCIDAR~\cite{Shepherd2014, Osborn2013, Osborn2017}.
Note that triangulation results in a strong and simple geometrical constraint (see (iv)) that facilitates profiling. However, the probed turbulence volume is limited to distances below $z_{max} \approx D/\Theta$, where $D$ represents the pupil diameter. 
A third strategy, which provides an alternative to triangulation, is to perform single source scintillation measurements and use the dependence with $z$ of the scintillation signature (iii) to perform the profiling, a feature exploited by MASS \cite{Tokovinin2003, Kornilov2003}, by SHABAR \cite{Beckers1999}, and by Single Star SCIDAR \cite{Habib2006}. However, relying on this sole information leads to a limited spatial resolution.

This is why we have developed a fourth strategy called CO-SLIDAR \revision{(COupled SLope and scIntillation Detection And Ranging)}~\cite{Vedrenne2007, Voyez-a-14}, which has the advantage of using all physical properties (i) to (iv) thanks to the measurement of both slope and scintillatin on a double source. In practice, all these measurements are deduced from Shack-Hartmann data. 
We again stress the fact that even if triangulation, on slopes and scintillation, is a strong asset for profiling below $z_{max}$, profiling is also permitted beyond $z_{max}$ thanks to the scintillation signature (see (iii)).
CO-SLIDAR has been applied to vertical profiling using a double star~\cite{Voyez-a-14} and more recently  \revision{profiling with near ground line of sight} with an instrument called Scindar \cite{Robert2015, Nguyen2017,nguyen_mesures_2018}. This instrument operates over a few kilometer range and uses two cooperative sources.  Detection is performed in the infra-red~\cite{Robert2012} to limit the turbulence optical effects so as to stay in the weak perturbation Rytov regime. Operation with a single source has also been demonstrated \cite{Vedrenne2010} proving the ability to exploit scintillation signature and therefore to bypass the maximum distance $z_{max}$ imposed by triangulation. Going beyond $z_{max}$ is an essential feature for near ground profiling where one has to probe the whole volume between pupil and sources. This limitation is a clear drawback for near ground profiling with SCIDAR/G-SCIDAR techniques that are based on triangulation on a double source \cite{Johnston2003}.

The CO-SLIDAR technique, which exploits both slopes and scintillation, is therefore very promising for the profiling of turbulence characteristics in the surface layer with a single instrument. Preliminary results have been published in \cite{Robert2015, Nguyen2017,nguyen_mesures_2018}, however the formalism has since been further developed (corrections made, several approximations alleviated). Besides the regularization approach has been improved, and the data processing and profile reconstructions fully updated accordingly. The present paper therefore presents the first comprehensive description of the most up-to-date CO-SLIDAR method and of its application to experimental data in the context of near ground metrology.

This paper includes the presentation of the physics principles underlying the measurements, which accounts for the specific geometry: finite distance propagation hence spherical waves, proper aperture and source filtering. We also present our reconstruction strategy: a Maximum \emph{A Posteriori} solution with a white quadratic regularization. An unsupervised regularization parameter adjustment based on the minimization of the generalized cross validation (GCV)~\cite{golub_generalized_1979} function is proposed, validated and used. 
Moreover, we estimate error bars on the $C_n^2$ reconstructed values. The Scindar instrument is presented and our unsupervised reconstruction strategy is applied to data acquired in a heterogeneous rural landscape during an experimental campaign in Lannemezan (France). The retrieved $C_n^2$ profiles are presented. The $C_n^2$ dependence in space and time is discussed. Results are also compared with scintillometer measurements. 

This paper is organized as follows. Section \ref{lannemezan} presents the Scindar instrument and the experimental campaign in Lannemezan. Section~\ref{coslidar} presents the CO-SLIDAR method in the near ground context (formalism and reconstruction algorithm). Section \ref{data-check} is dedicated to the data quality check for the unsupervised $C_n^2$ profile reconstruction. 
Section \ref{Cn2_profile} presents and discusses the reconstructed profiles, and their variability, observed in the heterogeneous rural landscape at Lannemezan. Conclusions and perspectives are outlined in Section \ref{conclusion}. The Appendix is dedicaded to the validation of the GCV method.

\section{Near ground horizontal 
\texorpdfstring{$C_n^2$}{Cn2} profiling campaign} \label{lannemezan} 
\begin{figure}[!htb]
    \centering
    	\begin{center}
		\begin{tabular}{cc}
		\includegraphics[width=0.55\textwidth]{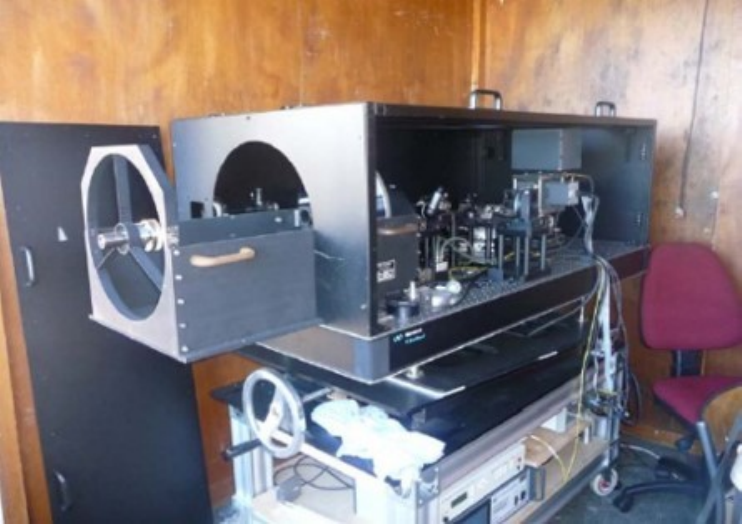} & \includegraphics[width=0.37\textwidth]{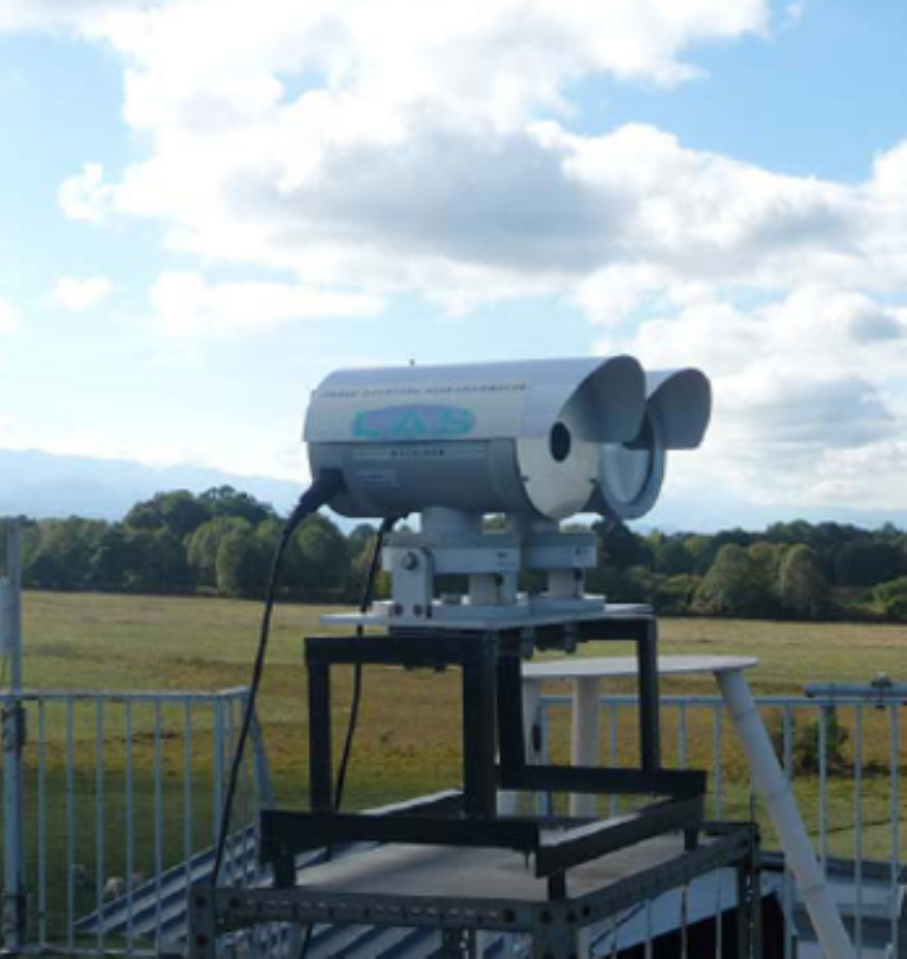} \\
		(a) & (b) \\
		 \multicolumn{2}{c}{\includegraphics[width=0.95\textwidth]{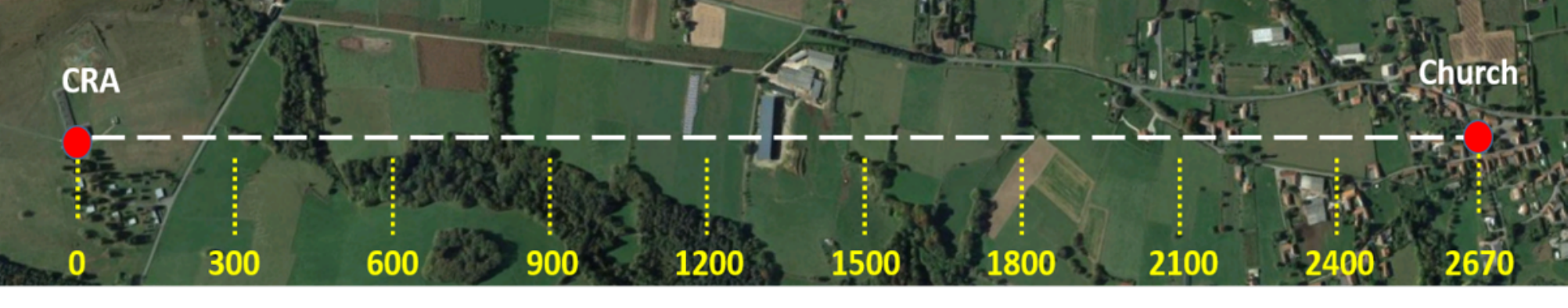}} \\
		 \multicolumn{2}{c}{(c)} 
		\end{tabular}
	\end{center}
    \caption{Experimental rural scintillometry campaign at the CRA of Lannemezan. (a) The Scindar Shack-Hartmann mid-infrared $C_n^2$ profiler at CRA, (b) LAS Kipp$\&$Zonen scintillometers at CRA, (c) Mosaic of heterogeneous rural covers below the line of sight.}
    \label{figure1}
\end{figure}

The consortium made of  ONERA, ISPA-INRAE and IGE (formerly LTHE) has conducted, in the late summer 2012, a field campaign to perform the first demonstration of a new $C_n^2$ profiler --~the Scindar~-- able to document the $C_n^2$ variability in the atmospheric surface layer. In this campaign (see Fig.~\ref{figure1}), additionally to the Scindar, three co-aligned scintillometers were also deployed.

\begin{figure} [!htb]
	\begin{center}
		\begin{tabular}{c}
			\includegraphics[trim = 2.6cm 1.1cm 2.3cm 2.2cm,clip,width=1\textwidth]{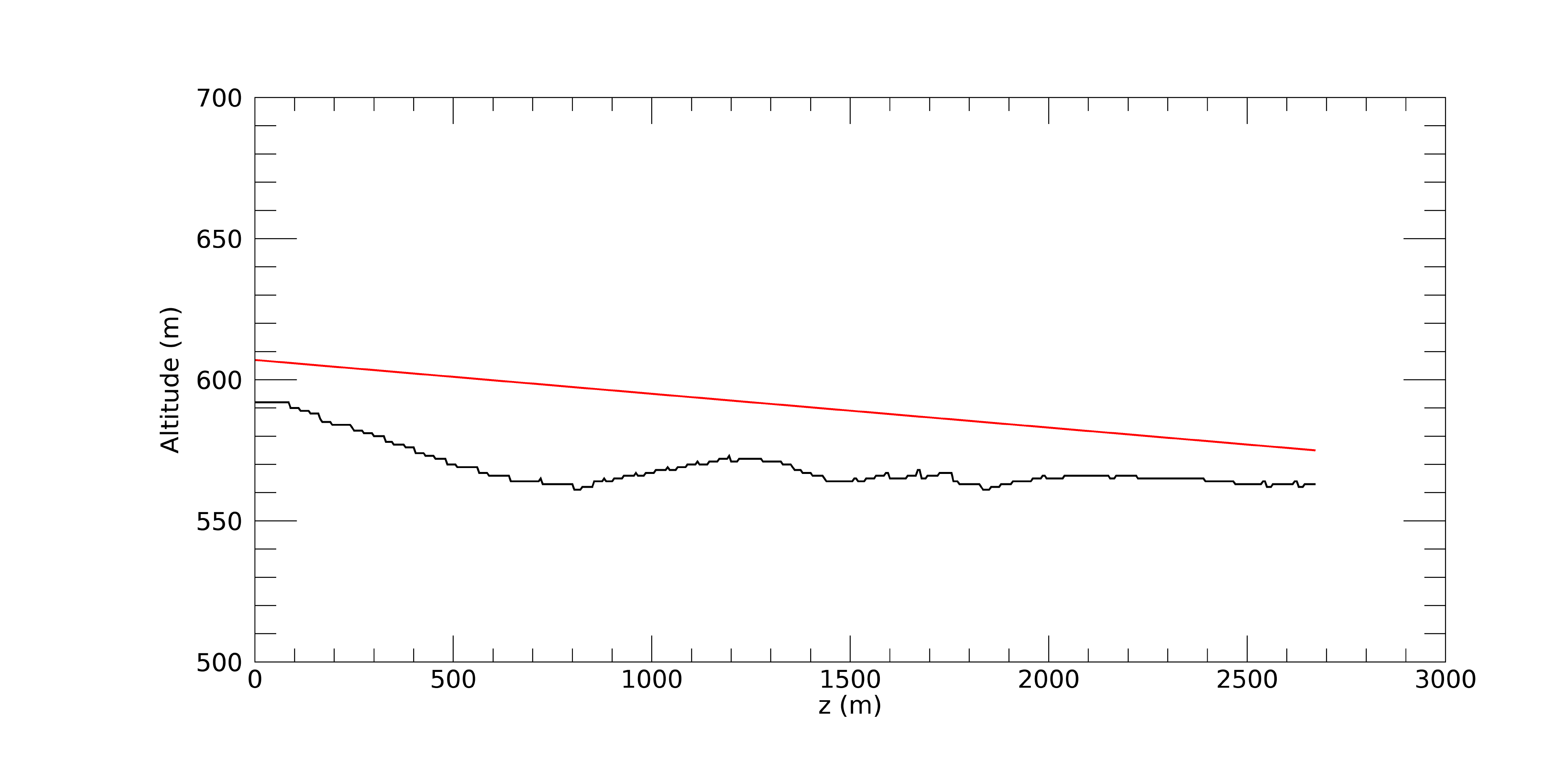}
		\end{tabular}
	\end{center}
	\caption {The topography of the line of sight as a function of $z$, i.e. the distance from the CRA. The terrain profile is plotted in black line and the line of sight in red line.}
	\label{fig:f11bis}
\end{figure}

The Scindar and the scintillometers were installed in parallel along a $L$ = 2670\,m line of sight at an average height of 23\,m above ground, between the CRA (Centre de Recherche Atmosphérique) facilities near Lannemezan ($43^o07^\prime 42.6^{\prime\prime}$ N, $0^o22^\prime02.1^{\prime\prime}$ E) and the church of the nearby Campistrous village ($43^o09^\prime 06.4^{\prime\prime}$ N, $0^o22^\prime35.4^{\prime\prime}$ E).
Figure~\ref{fig:f11bis} displays the topography of the  line of sight versus the distance to CRA denoted $z$.
Below the instrument beams, the landscape shows heterogeneities including grassland, tree lines and sparse houses of Campistrous. 
Of course the $C_n^2$ at a given position along the line of sight is influenced by many parameters: the height above ground, the topography, the ground nature (cover and surface roughness), the local temperature gradients, etc. The Scindar is dedicated to the $C_n^2$ metrology along the line of sight.

The detailed location and aperture characteristics of the available instruments are summarized in Table~\ref{table:t1}. The Scindar profiler consists of a Shack-Hartmann WaveFront Sensor (SHWFS) operating at mid-infrared (3.4--4.2\,$\mu$m) that observes two cooperative light sources. The SHWFS has 5$\times$5 square subapertures of size $d=7$\,cm and its images are recorded at a 142\,Hz frame rate. In this experiment, the SHWFS sensor aims at a double halogen source, 0.8\,m apart in the vertical direction ($y$) and whose front glasses (16$\times$12.5\,cm) act as thermal sources heated by the filament.In practice, the Shack-Hartmann wavefront sensor saves a set of $20$ effective subaperture images of the double source.

\begin{table}[!htb]
\begin{center}
\caption{Instruments location with their characteristics.}	
	\label{table:t1}
\begin{tabular}{|c|c|c|c|}
\hline
(d = diameter)& At CRA & At church & wavelength $\lambda$ \\
\hline \hline
\multirow{3}{*}{Scindar} &\multirow{2}{*}{mid-IR SHWFS} &  two halogen sources &\multirow{3}{*}{3.4 - 4.2 $\mu$m} \\
 & \multirow{2}{*}{D=$0.35$ m}  &(size $16 \times 12.5$~cm &  \\
 & & and $0.8$~m separation) & \\
\hline
Scintillometer A & $D_r$ = 5\,cm &   $D_t$ = 14.5\,cm & \multirow{2}{*}{0.88 $\mu$m} \\
\cline{1-3}
Scintillometer B & $D_r$ = 14.5\,cm & $D_t$ = 14.5\,cm  &  \\
\hline
Scintillometer C &  $D_t$ = 14.5\,cm &  $D_r$ = 5\,cm   & 0.94 $\mu$m\\
\hline	
\end{tabular}
\end{center}
 
\end{table}

\revision{Three scintillometers have been installed in parallel to the Scindar line of sight: the two commercial Kipp$\&$Zonen brought by LAS, shown on Fig.~\ref{figure1}b  and denoted A and B, were located at CRA, and a prototype built at Wageningen University, denoted C, was located in the church steeple. The use of different aperture diameters at the transmitter ($D_t$) and at the receiver ($D_r$) side allows one to obtain complementary sensitivities on the line of sight: regarless of the propagation direction, the sensitivity is shifted towards the location of the smallest aperture~\cite{Wang1978}. In practice, simple circular diaphragms reducing the original aperture have been placed on the receivers of the two scintillometers A and C (see Table~\ref{table:t1}). The sensitivity of the three scintillometers is given in Sect.~\ref{Cn2_profile}.}

In the following, we exploit the data acquired in the afternoon of September $14^{th}$, 2012. Surface conditions were moist due to a small rainy event the night before. The Scindar data have been processed for two 100 minutes periods (in local time): from 14:15 to 15:55  when stationary turbulence conditions have been observed; from 17:15 to 18:55  when declining turbulence  \revision{has} been observed because of the decaying solar radiation. This was a sunny day, note that the sunrise was at 7:37, the sunset at 20:10 and the sun culmination was at 13:53 (local time).

\section{The CO-SLIDAR method: case of near ground horizontal profiling}
\label{coslidar}
This section describes the CO-SLIDAR method in the context of near ground $C_n^2$ profiling \revision{over} a finite distance. We introduce in Subsection~\ref{discretization} the observation geometry and the notion of triangulation in the Scindar configuration. We define in Subsection~\ref{formalism} the slope and scintillation correlations that are derived from the Shack-Hartmann measurements and we give a physical description that relates the $C_n^2$ profile to these correlations.   
We define in Subsection~\ref{direct_model} the reduced data model associated to the profile discretization. Our reconstruction strategy is, finally, presented in Subsection~\ref{regularization}: a Maximum \emph{A Posteriori} (MAP) solution using a white quadratic regularization with an unsupervised hyper-parameter adjustment.

\subsection{CO-SLIDAR triangulation} \label{discretization}

\begin{figure} [!htb]
	\begin{center}
		\begin{tabular}{c}
			\includegraphics[trim = 1cm 3cm 0.6cm 3cm,clip,width=1\textwidth]{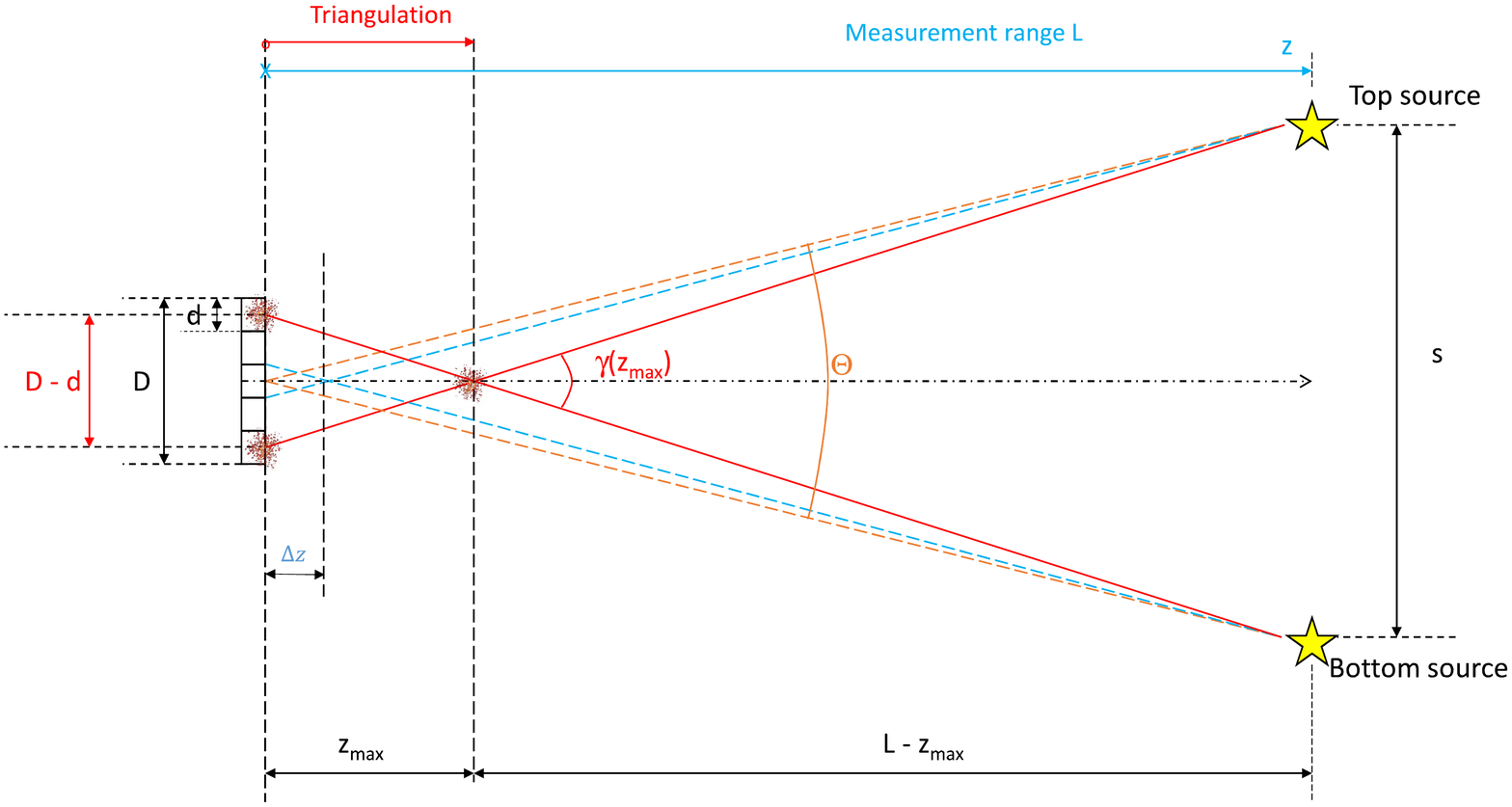}
		\end{tabular}
	\end{center}
\caption{CO-SLIDAR principle: multi-aperture observation of a double source ($z=0$ at SHWFS plane, $z=L$ at source plane).}
	\label{fig:f2}
\end{figure}

The observation geometry is shown on the Fig.~\ref{fig:f2} in cross section. The SHWFS on the left is  constituted of $p \times p$ subapertures, illustrated in the Scindar configuration (here $p = 5$). We denote by $d$ the subaperture diameter while $D$ is the SHWFS overall pupil diameter. $L$ is the distance between the SHWFS pupil and the source plane. 
$s$~is the vertical distance between the two sources and $\Theta = s/L$ is the source angular separation as seen from the pupil. It can be easily shown that the complex amplitudes of the electromagnetic field, observed in the SHWFS pupil, and induced by a given \revision{slice} at $z$ are identical for the two sources apart from a translation given by:
\begin{equation}
\label{translation}
    \gamma(z) \ z =  \frac{s}{(L-z)} \ z  
    \end{equation}
where $\gamma(z)$ is the source \revision{angular} separation as seen from a slice at z. Triangulation exploits this fundamental property to retrieve the turbulence distribution along the line of sight. Triangulation, however, operates in a limited range.

The maximum distance of triangulation $z_{max}$ corresponds to the distance $z$ for which the turbulent slice effects (slope and scintillation) perceived in the SHWFS pupil plane shift by $(p-1)$ subapertures (\textit{i.e} a distance D - d) when going from one source to the other. One can therefore write:
\begin{equation}
\gamma(z_{max}) = \frac{s}{L-z_{max}} =\frac{D-d}{z_{max}} = \frac{D-d+s}{L}.
\end{equation}
Hence a maximum distance of triangulation $z_{max}$ given by:
\begin{equation}
z_{max}=\frac{D-d}{D-d+s}L.
\label{eq:5}
\end{equation}
We obtain $z_{max}= 692\, m$ for the parameters of the Scindar observation.

Concerning the triangulation longitudinal resolution it is often deduced from the following simple geometrical reasoning. We can consider that the first slice that can be distinguished from the pupil slice is located at $z=\Delta z$, if slope and scintillation  patterns shift by one subaperture when going from one source to the other. By using Thales' theorem and an arithmetical rule on the quotients, one can therefore write:
\begin{equation}
\gamma(\Delta z) = \frac{d}{\Delta z} = \Theta \frac{L}{L-\Delta z} = \frac{d+\Theta L}{L}.
\end{equation}
Hence a resolution $\Delta z$ given by:
\begin{equation}
\Delta z = \frac{d}{d+\Theta L}L = \frac{d}{d+s}L \approx d\frac{L}{s} = d \Theta.
\end{equation}
We obtain here $\Delta z \approx 214\, m$. In the rest of the paper we take a uniform sampling of the $C_n^2$ profile with a number of turbulent slices along the line of sight $k =\frac{L}{\Delta z}$ rounded to $k=12$, then corresponding to a sampling step $\Delta z = 223 \, m$.

Our \revision{reasoning up} to now concentrated on the sole triangulation feature of the CO-SLIDAR method. However, we show in the next section that this method does not solely rely on triangulation and actually allows $C_n^2$ profiling on the whole line of sight, from the SHWFS pupil to the source plane. This is an essential feature for our near ground application and is a usual asset of the profilers that record scintillation patterns, as explained in the introduction.

\subsection{Slope and scintillation correlations: triangulation and beyond} \label{formalism}
At a given time $t$, for each source image, we compute two slopes and one scintillation index per subaperture. For a given source identified by its angular position $\alpha$, the slopes measured in the subaperture $a$ is  denoted $s_a(\alpha)$. This bi-scalar vector contains the slopes along x and y: $s_a^l(\alpha)$, $l \in \{x,y\}$. It is computed as the center of gravity of the subimage in a moving window centered on the maximum of source image. 
The scintillation index $\delta i_a(\alpha)$ is computed from the source intensity detected in subaperture $a$ denoted $i_a(\alpha)$ as follows: $\delta i_a(\alpha) =\frac{i_a(\alpha)-\left< i_a(\alpha) \right>}{\langle i_a(\alpha)\rangle}$ where $\left< \right>$ represents temporal average over a finite time series. The intensity $i_a(\alpha)$ is the sum of the pixel intensities in the moving window.

The essential quantities for CO-SLIDAR are the slope and scintillation correlations defined respectively as $\left< s_a^l(\alpha)s_b^m(\alpha+\theta) \right>$ and $\left<\delta i_a(\alpha) \delta i_b(\alpha+\theta) \right>$, where $\theta \in \{0,\Theta\}$. $\theta=0$ corresponds to the so-called auto-correlation (correlation of the slope and scintillation data for a given source),  while $\theta=\Theta$  corresponds to the so-called cross-correlation (correlation of the slope and scintillation data from two different sources).
 
Assuming small perturbations in the Rytov regime, \revision{and independence of the phase perturbations between slices,}  the slope and scintillation correlations \revision{between subapertures $a$ and $b$ can be expressed as the sum of contributions associated to each individual slice. Besides, the contribution of a slice, at distance $z$ from the pupil and of thickness $dz$, can be shown~\cite{Robert-a-06} to be proportional to the local turbulence strength, hence to $C_n^2(z)\,dz$. One can then express~\cite{Robert-a-06, Vedrenne2010} these correlations as simple integral equations with a linear dependence in $C_n^2(z)$ that takes the following form:}
\begin{equation}
\label{eq:7}
\left< s_a^l(\alpha)s_b^m(\alpha+\theta) \right> =\int_{0}^{L}W_{ss}^{lm}(d_{ab},\theta,z)C_n^2(z)dz,
\end{equation}
\begin{equation}
\label{eq:8}
\left<\delta i_a(\alpha) \delta i_b(\alpha+\theta) \right> =\int_{0}^{L}W_{ii}(d_{ab},\theta,z)C_n^2(z)dz,
\end{equation}
where $(l,m) \in \{x,y\}$\revision{, $d_{ab}$ is the separation vector between the considered subapertures, and the response functions $W$ are the normalized contributions, for a slice at $z$, and for $C_n^2(z) dz = 1$.}
\revision{In the CO-SLIDAR method, we only use the correlations of $x$-slopes, of $y$-slopes  and of course of scintillation. 
We do not use inter-correlations between $x$ and $y$ slopes or between slopes and scintillation. We do so because we observed in other contexts that the use of these inter-correlations had little impact on the reconstructed profiles.}

\revision{The response functions $W$, that is to say the normalized single layer correlations as a function of $d_{ab}$, can be easily expressed in the Fourier domain \emph{i.e.}, in terms of their respective spatial power spectral density functions defined as follows:}
\begin{equation}
\label{eq:9}
W_{ss}^{xx}(d_{ab},\theta,z)=FT^{-1}[G_{ss}^{xx}(\boldsymbol{f},\theta,z)](d_{ab}),
\end{equation}
\begin{equation}
\label{eq:10}
W_{ss}^{yy}(d_{ab},\theta,z)=FT^{-1}[G_{ss}^{yy}(\boldsymbol{f},\theta,z)](d_{ab}),
\end{equation}
\begin{equation}
W_{ii}(d_{ab},\theta,z)=FT^{-1}[F_{ii}(\boldsymbol{f},\theta,z)](d_{ab}),
\end{equation}
where $FT^{-1}$ represents the inverse Fourier Transform with respect to the spatial frequency $\boldsymbol{f}=(f_x, f_y)$, which is dual of the spatial variable $d_{ab}$.  
\revision{The power spectral density functions $G_{ss}^{ll}$ and $F_{ii}$ account for the beam geometry, for the pupil and source filtering effects and for the Fresnel propagation between a given slice and the aperture. Their expressions can be found in \cite{Robert-a-06, Vedrenne2010} in the case of point sources at infinity,  called plane wave scenario.} We present hereafter a generalization of these expressions in the case of extended sources at a finite distance. This spherical wave scenario indeed corresponds to the Scindar configuration. $G_{ss}^{ll}$ and $F_{ii}$ then take the following form:
\begin{equation}
\label{eq:11}
\begin{split}
G_{ss}^{xx}(\boldsymbol{f},\theta,z)=(2\pi)^2f_x^2 \left[ S_n^z\left(\frac{L}{L-z} \boldsymbol{f}\right) \right] \left(\frac{L}{L-z}\right)^{2}\cos^{2}\left(\pi\lambda\frac{L}{L-z}z||\boldsymbol{f}||^2\right) \times \\ F_{pupil}(\boldsymbol{f}) F_{source}(\boldsymbol{f}) \exp\left(-2i\pi z\theta \left(\frac{L}{L-z} \right) f_y \right) ,    
\end{split}
\end{equation}

\begin{equation}
\begin{split}
G_{ss}^{yy}(\boldsymbol{f},\theta,z)=(2\pi)^2f_y^2 \left[ S_n^z\left(\frac{L}{L-z} \boldsymbol{f}\right)\right] \left(\frac{L}{L-z}\right)^{2}\cos^{2}\left(\pi\lambda\frac{L}{L-z}z||\boldsymbol{f}||^2\right) \times \\ F_{pupil}(\boldsymbol{f}) F_{source}(\boldsymbol{f}) \exp\left(-2i\pi z\theta \left(\frac{L}{L-z} \right) f_y \right) ,  
\end{split}
\end{equation}

\begin{equation}
\label{Eq:13}
\begin{split}
F_{ii}(\boldsymbol{f},\theta,z)=4\left(\frac{2\pi}{\lambda}\right)^2  \left[S_n^z\left(\frac{L}{L-z} \boldsymbol{f}\right)\right] \left(\frac{L}{L-z}\right)^{2}\sin^{2}\left(\pi\lambda\frac{L}{L-z}z||\boldsymbol{f}||^2\right) \times \\ F_{pupil}(\boldsymbol{f}) F_{source}(\boldsymbol{f}) \exp\left(-2i\pi z\theta \left(\frac{L}{L-z} \right) f_y \right),    
\end{split}
\end{equation}
where $\lambda$ is the wavelength, $F_{pupil}(\boldsymbol{f})$ and $F_{source}(\boldsymbol{f})$ are the spatial filters induced respectively by pupil and source averaging detailed below, $S_n^z(\boldsymbol{f})$ \revision{is the power spectral density of the refractive index $n$ associated for a unitary $C_n^2(z) dz$. A Kolmogorov model is considered in the present paper, it therefore reads:}
\begin{equation}
\label{VKspectrum}
	\revision{S_n^z(\boldsymbol{f})= 0.033 (2\pi)^{-\frac{2}{3}} ||\boldsymbol{f}||^{-\frac{11}{3}}.}
\end{equation}

In Eqs.~(\ref{eq:11}-\ref{Eq:13}), the factor $\frac{L}{L-z}$ comes for the spherical wave propagation. The Fresnel propagation terms depend on the nature of the correlated measurements: $\cos^{2}\left(\pi\lambda\frac{L}{L-z}z||\boldsymbol{f}||^2\right)$ for correlations of slopes and $\sin^{2}\left(\pi\lambda\frac{L}{L-z}z||\boldsymbol{f}||^2\right)$ for those of scintillation.  
The term of phase shift $\exp(-2i\pi z\theta \left(\frac{L}{L-z}\right) f_y)$ is due to the separation between the sources (assumed along the y axis). Note that $\theta \left(\frac{L}{L-z}\right)$ is the source angular separation as seen from the slice at $z$  denoted $\gamma(z)$ in Eq.~(\ref{translation}). These terms and scaling factors are typical of Fresnel propagation with spherical waves \cite{Sasiela2007}.

The spatial filter associated to pupil averaging $F_{pupil}(\boldsymbol{f})$ is simply the square modulus of the Fourier transform of the aperture support. Here for the averaging by a square subaperture of side $d$: 
\begin{equation}
    F_{pupil}(\boldsymbol{f}) = \left|\tilde{P}_{pupil}(\boldsymbol{f})\right|^2= \mathrm{sinc}^2(\pi f_xd) \mathrm{sinc}^2(\pi f_yd)
\end{equation}

The spatial filter associated to source averaging is similarly related to square modulus of the Fourier transform of the source intensity distribution, with a frequency scaling accounting for spherical waves and for the fact that the filter is expressed in the SHWFS pupil plane. It eventually reads \cite{Sasiela2007}:
\begin{equation}
    F_{source}(\boldsymbol{f}) = \left|\tilde{P}_{source}\left(\frac{z}{L-z}\boldsymbol{f}\right)\right|^2 .
\end{equation}

We assume here that the source intensity distribution is described by a 2D Gaussian function with separable variables of the type: \revision{${P}_{source}(x,y)\,=\,\exp(-x^2/\beta_x^2) \exp(-y^2/\beta_y^2)$}, hence the Fourier transform: \revision{$\tilde{P}_{source}(\boldsymbol{f}) = \exp $ $  (- \pi^2  \beta_x^2 f_x^2)  \exp(- \pi^2 \beta_y^2 f_y^2)$}. The source size can indeed be fitted by a separable variable Gaussian function with Full Width at Half Maximum (FWHM) estimated at $8.9$\,cm along the $x$ axis and $6.3$\,cm along the $y$ axis. These values are used to compute $\beta_x$ and $\beta_y$.

\subsection{Reduced data model} \label{direct_model}
We precise here the definition of the reduced data and we express the reduced data model used in the MAP reconstruction.

Since the slope and scintillation correlations described in Sect.~\ref{formalism} depend on $d_{ab}$, the separation vector between two subapertures $a$ and $b$, we define the reduced data as the correlation maps obtained by averaging correlations over all pairs of subapertures with given separation vectors $d_{ab}$, hence three kinds of reduced data defined as:
\begin{equation}
\label{eq:16}
C_{ss}^{xx}(d_{ab},\theta)=\frac{\sum_{\{a,b\}_{d_{ab}}}\left< s_a^x(\alpha)s_b^x(\alpha+\theta) \right>}{N(d_{ab})},
\end{equation}

\begin{equation}
\label{eq:17}
C_{ss}^{yy}(d_{ab},\theta)=\frac{\sum_{\{a,b\}_{d_{ab}}}\left< s_a^y(\alpha)s_b^y(\alpha+\theta) \right>}{N(d_{ab})},
\end{equation}

\begin{equation}
\label{eq:18}
C_{ii}(d_{ab},\theta)=\frac{\sum_{\{a,b\}_{d_{ab}}}\left<\delta i_a(\alpha) \delta i_b(\alpha+\theta) \right>}{N(d_{ab})},
\end{equation}
where $\sum_{\{a,b\}_{d_{ab}}}$ denotes the summation over all pairs of subapertures with separation vector $d_{ab}$ and $N(d_{ab})$ represents the number of such pairs. The correlation map dimensions are $(2p -1) \times (2p -1)$ where $p$ is the number of sub-apertures across the SHWFS diameter.
The values obtained for each map can then be piled up in lexicographic order and concatenated in a single reduced data vector $C_{mes}$ of $6 (2p -1)^2$ elements:

\begin{equation}
C_{mes}=
\begin{pmatrix}
C_{ss}^{xx}(d_{ab},0) \\
C_{ss}^{yy}(d_{ab},0) \\
C_{ss}^{xx}(d_{ab},\Theta) \\
C_{ss}^{yy}(d_{ab},\Theta) \\
C_{ii}(d_{ab},0)  \\
C_{ii}(d_{ab},\Theta)
\end{pmatrix}.
\end{equation}
In practice $C_{mes}$ consists of correlations that are estimated from a finite number of noisy slope and scintillation data, therefore with biases induced by detection noise and convergence noise induced by the finite number of data. The measurement equation relating the $C_n^2$ profile to the reduced data can therefore be expressed in the following matrix form:
\begin{equation} \label{eq-modele-direct}
C_{mes} = M C_n^2 + C_d + u,
\end{equation}
where $C_d$ represents the slope and scintillation noise covariance matrix, which biases the reduced data estimation, while $u$ represents the convergence noise related to the finite number of data, which we assume to be Gaussian in the following. The matrix $M$ is derived from the discretization of Eqs.~(\ref{eq:7}) and~(\ref{eq:8}) and composed of the slopes and scintillation response functions to a unitary $C_n^2(z)$:
\begin{equation}
M=
\begin{pmatrix}
	W_{ss}^{xx}(d_{ab},0,z_1) \Delta z_1 \ ...\ W_{ss}^{xx}(d_{ab},0,z_i) \Delta z_i\ ...\ W_{ss}^{xx}(d_{ab},0,z_k) \Delta z_k\\
	W_{ss}^{yy}(d_{ab},0,z_1)  \Delta z_1\ ...\ W_{ss}^{yy}(d_{ab},0,z_i) \Delta z_i\ ...\ W_{ss}^{yy}(d_{ab},0,z_k,) \Delta z_k\\
	W_{ss}^{xx}(d_{ab},\Theta,z_1)  \Delta z_1\ ...\   W_{ss}^{xx}(d_{ab},\Theta,z_i) \Delta z_i\ ...\   W_{ss}^{xx}(d_{ab},\Theta,z_k) \Delta z_k\\
	W_{ss}^{yy}(d_{ab},\Theta,z_1)  \Delta z_1\ ...\   W_{ss}^{yy}(d_{ab},\Theta,z_i) \Delta z_i\ ...\   W_{ss}^{yy}(d_{ab},\Theta,z_k) \Delta z_k\\
	W_{ii}(d_{ab},0,z_1)  \Delta z_1\ ...\ W_{ii}(d_{ab},0,z_i) \Delta z_i\ ...\ W_{ii}(d_{ab},0,z_k) \Delta z_k\\
	W_{ii}(d_{ab},\Theta,z_1)  \Delta z_1\ ...\ W_{ii}(d_{ab},\Theta,z_i) \Delta z_i\ ...\ W_{ii}(d_{ab},\Theta,z_k) \Delta z_k
\end{pmatrix},
\label{eq:M}
\end{equation}
where we assumed that the turbulent volume is discretized in k \revision{slices} distributed along the line of sight at distances $z_i$, taking $z_0$ = 0, and with thicknesses $\Delta z_i$ ($i=1,2,...,k$). We recall that, as stated in Section~\ref{discretization}, we consider k=12 equidistant slices of thickness $\Delta z_i = \Delta z = 223 $~m.

\newpage
	\begin{figure}[!htb]
     \centering
    	\begin{center}
		\begin{tabular}{c}
		\includegraphics[trim = 4.3cm 1.4cm 3.cm 1.5cm, clip,width=1\textwidth]{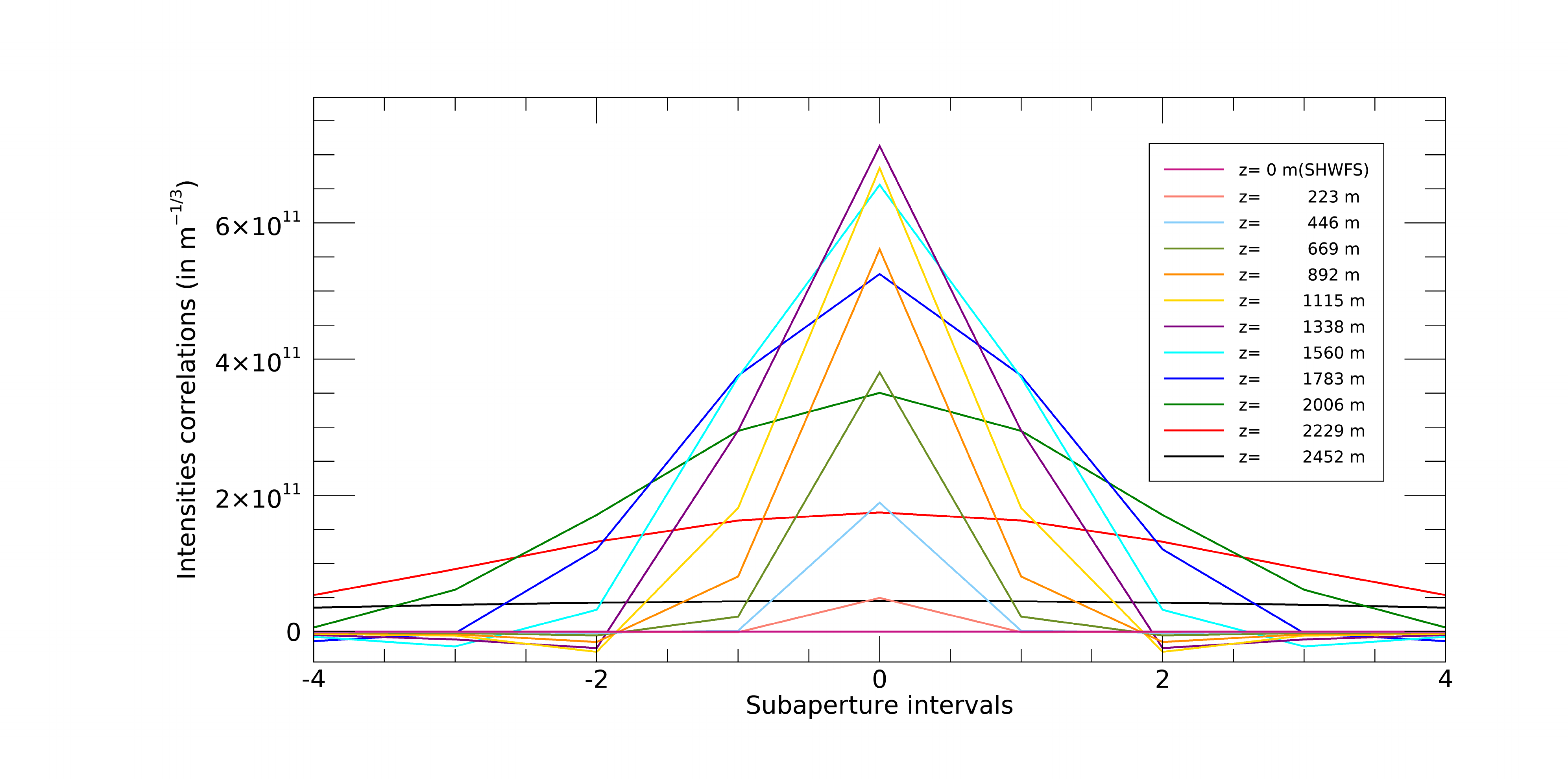} \\
	    	(a) \\
		    \includegraphics[trim = 3.8cm 1.4cm 3.cm 1.5cm, clip,width=1\textwidth]{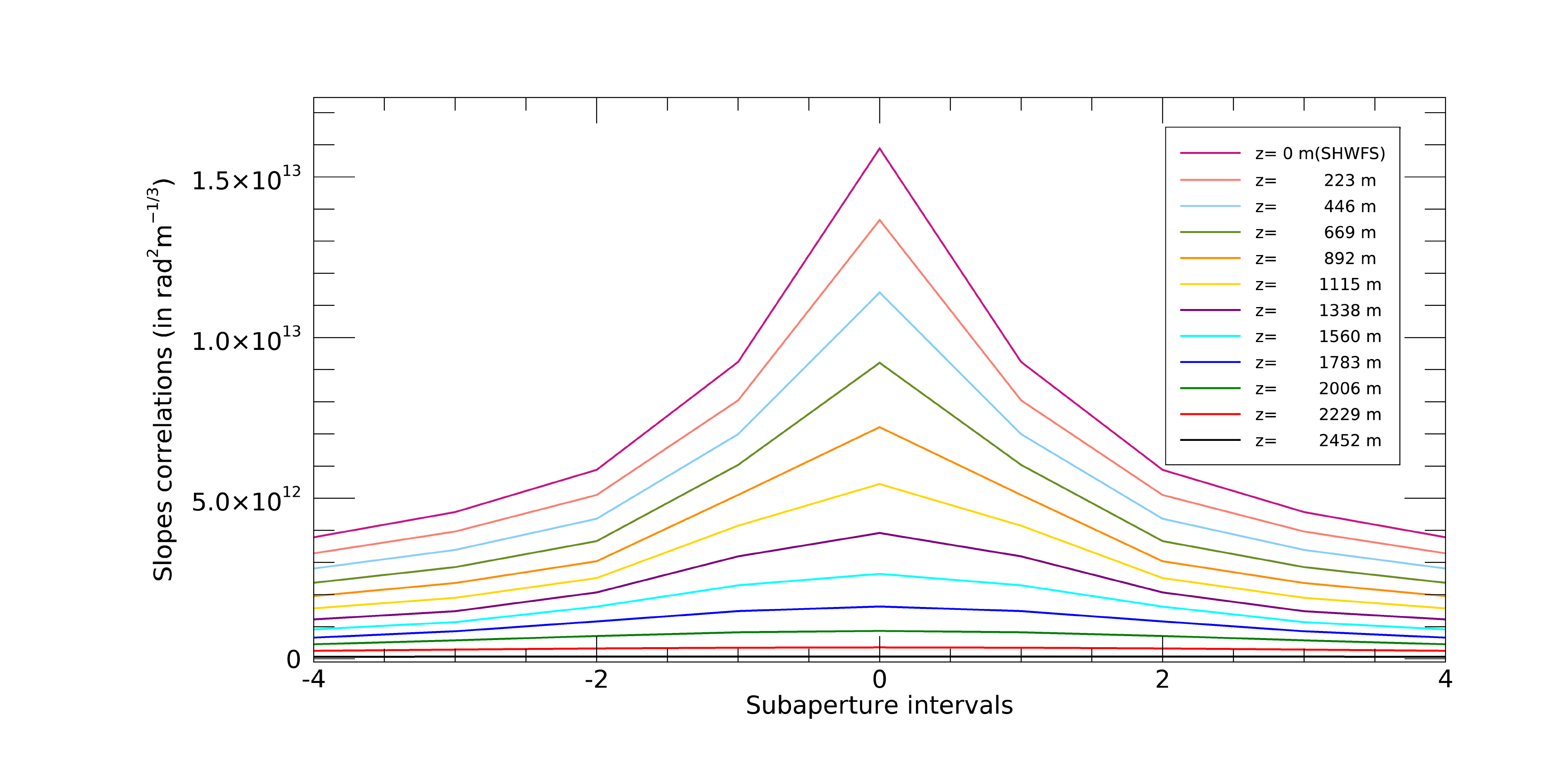} \\
	        (b) 
	        \end{tabular}
	\end{center}
	 \caption{Scindar scintillation and slopes auto-correlation response to a unitary $C_n^2(z) \Delta z$ for each of the 12 slices. (a) Intensities auto-correlation, (b) x slopes auto-correlation as a function of the subaperture intervals defined as the ratio $d_{ab}/d$. 	
	 }   
	 \label{fig:poid}
    
\end{figure}

	\begin{figure}[!htb]
    \centering
    	\begin{center}
		\begin{tabular}{c}
	 \includegraphics[trim = 4.3cm 1.4cm 3.cm 1.5cm, clip,width=1\textwidth]{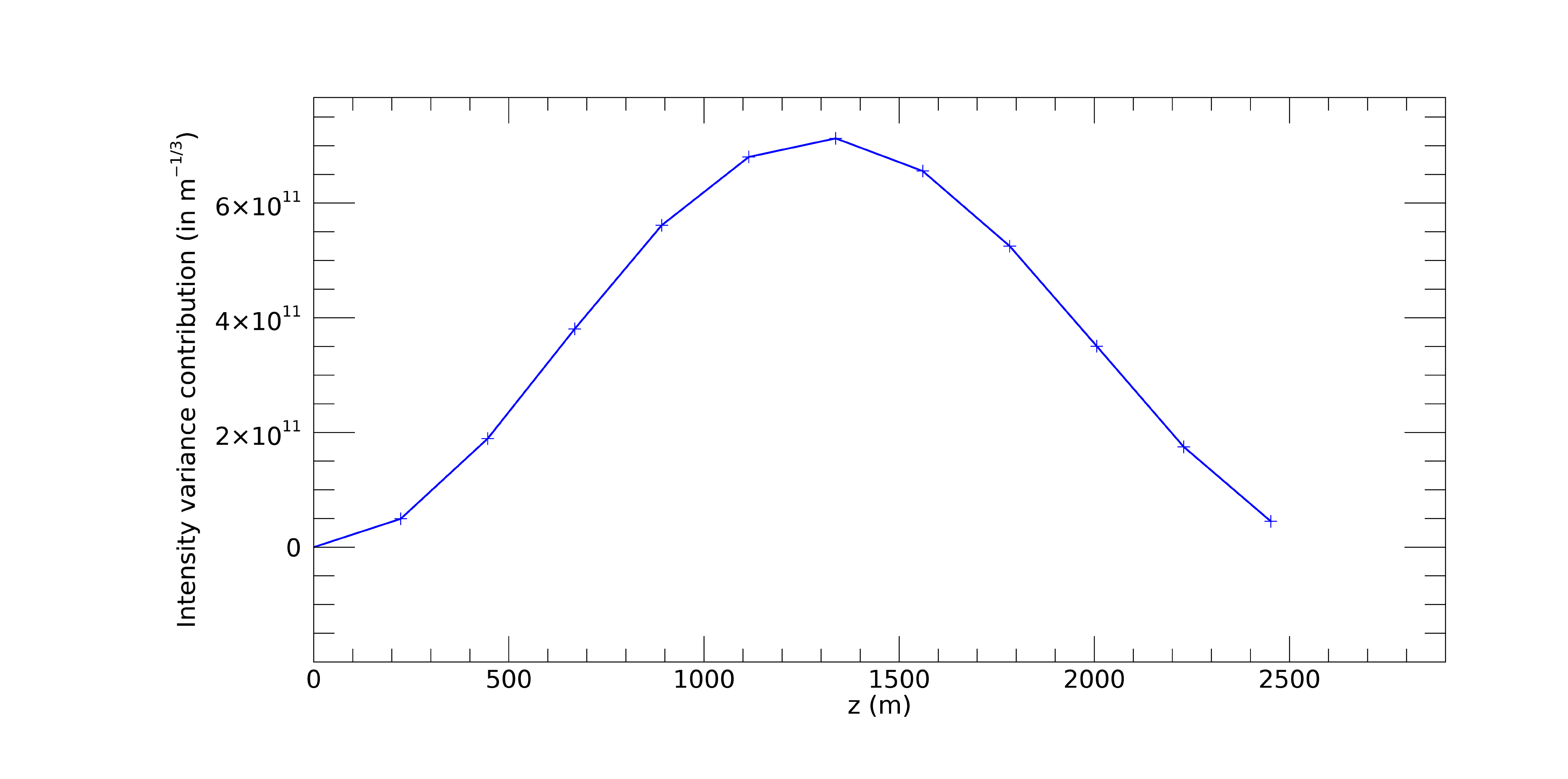} \\
	  
		(a) \\
		\includegraphics[trim = 3.8cm 1.4cm 3.cm 1.5cm, clip,width=1\textwidth]{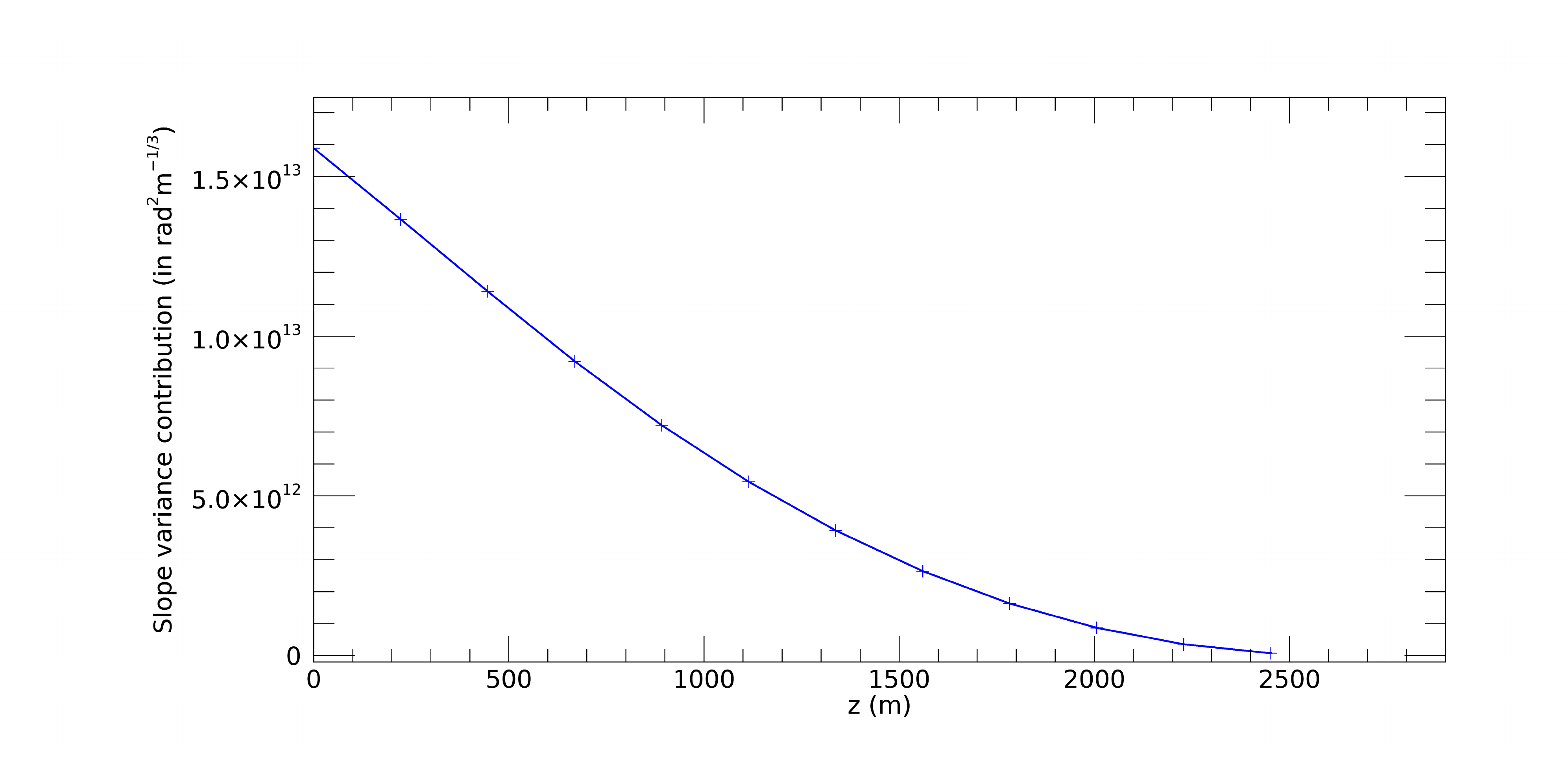} \\
	    (b) 
	        \end{tabular}
	\end{center}
 \caption{(a) Intensity (b) slope variances contribution as a function of the distance to SHWFS (receiver).
 }   
 \label{fig:f4}
\end{figure}

Figure~\ref{fig:poid} presents different scintillation and slope auto-correlation response functions defined in Eqs.~(\ref{eq:9}-\ref{eq:11}) by setting $\theta=0$. These responses depend on $d_{ab}$ for each of the 12 individual slices. The shape of the curves related to intensities (see Fig.~\ref{fig:poid}a) depends on the distance $z$. It can be easily derived from Eq.~(\ref{Eq:13})  that the scintillation pattern typical size, in spherical wave and for a given slice at z, is $\sqrt{\lambda z  L/(L-z)}$. A consequence of the dependence of this expression with $z$ is then observed on the Fig.~\ref{fig:poid}a where the scintillation's correlation width clearly varies with $z$, as opposed to the slope correlation curves (see Fig. \ref{fig:poid}b) which are very weakly dependent on $z$, apart from a scaling factor. This clear dependence in $z$ of the scintillation size, hence of the scintillation correlation structure, is a salient feature for the slice localization and in turns permits $C_n^2(z)$ profiling all along the line of sight.

On the Fig.~\ref{fig:f4}, the maximum of the  auto-correlation response functions, that is the slope and scintillation variances for $C_n^2(z_i) \Delta z_i = 1$, is also drawn as a function of $z$. 
The latter curves give an idea of the contribution to slopes and scintillation as a function of $z$. The slope contribution decreases monotonously with $z$ because of the spherical wave geometry (sources at finite distance) (see Fig.\ref{fig:f4}b). The scintillation contribution follows a bell curve (see Fig.\ref{fig:f4}a). Contribution to scintillation is of course null at $z=0$ (no diffraction effect), it then increases with $z$ as diffraction effects build up (Fresnel propagation) and decreases near the sources because of spherical waves. The scintillation brings a significative contribution close to the middle of the path. Note that this almost symmetrical bell-shape curve is expected since the estimated size of the source's FWHM and the subaperture sizes are similar, as for the weighting function of a scintillometer with symmetrical apertures (see Fig.~\ref{fig:fct_poid}).
Scindar exploits the sensitivity of slopes and scintillation to turbulence slices on the whole line of sight.

\subsection{Maximum \emph{A Posteriori} solution with white quadratic regularization} \label{regularization}

The reconstruction of the discretized $C_n^2$ profile is performed in a Bayesian framework, more precisely with the MAP or penalized maximum likelihood method---see, \textit{e.g.}, \cite{Mugnier-l-12,Idier-l-08} for tutorial material. The MAP approach boils down to minimizing a metric composed of two terms, a data fidelity term and a regularization term. The data fidelity term is the opposite of the log-likelihood, which under the Gaussian assumption adopted here for the convergence noise $u$ takes a quadratic form. The regularization term, denoted by $J_{\text{prior}}$ in the following, embodies our prior knowledge on the regularity of the sought profile and aims at penalizing the profiles that are incompatible with this prior knowledge. 
Using Eq.~(\ref{eq-modele-direct}), the reconstructed profile is thus the minimum of the following metric:
\begin{equation}
J_{MAP}\left(C_n^2\right)=\left(C_{mes}-C_d-MC_n^2\right)^TC_{conv}^{-1}\left(C_{mes}-C_d-MC_n^2\right)+J_{\text{prior}}\left(C_n^2\right),
\label{eq:JMAP}
\end{equation}
where $C_{conv}=\left<u u^T\right>$ is the covariance matrix of the convergence
noise u, and is deduced from an analytical expression depending on the theoretical
correlations, which are in practice approximated with the associated experimental correlations (see Appendix~A of Ref.~\cite{Voyez-a-14}). 
The detection noise covariance matrix  $C_d$ appears in Eq.~(\ref{eq:JMAP}) as a bias which is estimated jointly with the $C_n^2$ profile \cite{Voyez-a-14}.
Previous works have experimented with various regularization metrics: \cite{Vedrenne2007} used a metric based on a weighted Laplacian of the profile, with weights taking into account the expected strength of each slice, assuming a nighttime Hufnagel–Valley profile. \cite{Voyez-a-14} used a simpler metric based on the gradient of the profile. Both metrics are quadratic and promote correlated, smooth profiles. In this paper, because we address horizontal $C_n^2$ profiling where terrains, covers and weather conditions are quite complex, we choose a regularization that remains quadratic, which amounts to a Gaussian assumption on the prior probability for the sought profile, but is uncorrelated or white --~in the sense that it assumes independent slices.

In the Bayesian framework adopted here, the regularization metric $J_{\text{prior}}$ is the opposite of the logarithm of the prior probability, so it is given by:
\begin{equation} 
J_{\text{prior}}\left(C_n^2\right)=\sum_{i=1}^k \left(\frac{C_n^2(i)}{\sigma_{\text{prior}}(i)}\right)^2 ,
\label{eq:JPRIOR}
\end{equation}
where $\sigma_{\text{prior}}(i)$ is the \textit{a priori} variability of slice $i$ of the sought $C_n^2$ profile. For simplicity, we suppose that this variability is independent of the slice ($\sigma_{\text{prior}}(i)=\sigma_{\text{prior}} \,\forall i$), so the regularization metric can be rewritten as:
\begin{equation}\label{eq-Jprior-white-homog-quadratic}
J_{\text{prior}}\left(C_n^2\right)={\mu}\sum_{i=1}^k \left(C_n^2(i)\right)^2,
\end{equation}
where $\mu = 1/\sigma_{\text{prior}}^2$ is the only hyper-parameter of the metric to optimize. This hyper-parameter adjusts the balance between the likelihood term and the prior term.

The minimization of $J_{MAP}$ is done under positivity constraint so that the $C_n^2$ only takes positive values and it is performed with the Variable Metric with Limited Memory and Bounds (VMLM-B) algorithm \cite{Thiebaut2002}. 

To adjust the hyper-parameter of regularization we use the statistical method of the Generalized Cross Validation \revision{(GCV)~\cite{golub_generalized_1979,Demoment-l-08a}}.
The adaptation of the GCV function ($V(\mu)$) for our model and its evaluation on synthetic data is made on Appendix. By minimizing the GCV function we can obtain a reasonable value for the hyper-parameter.

Obtaining error bars on the reconstructed profile is an important feature for the user. 
\revision{In the current statistical framework where our estimator is the mode of the \emph{a posteriori} probability distribution  of the sought profile, error bars are usually estimated from the \emph{a posteriori} covariance matrix~\cite{efron_frequentist_2015}. More precisely, 1$\sigma$  error bars are computed as the square root of the diagonal values of the posterior covariance matrix, which can easily be shown to be:}
\begin{equation}\label{eq-error-covariance}
C_{error} = (M^T C_{conv}^{-1} M + \mu I)^{-1} ,
\end{equation}
where $I$ is the identity matrix. 
\revision{These error bars are somewhat conservative, due to the fact that they characterize the variability of the solution obtained without a positivity constraint. Indeed, the covariance matrix of Eq.~(\ref{eq-error-covariance}) does not take into account the stabilization of the solution brought by the positivity constraint of the reconstruction.}

A difficulty to obtain reliable error bars is that the error covariance matrix depends on the chosen prior and notably on the chosen value for $\mu$. In particular, if $\mu$ is chosen too large, the solution is substantially biased towards zero and the error bars become, in practice, smaller than the actual error on the profile. 
\revision{Conversely,} by setting  $\mu=0$ in Eq.~(\ref{eq-error-covariance}), one obtains
regularization-independent and (very) conservative error bars, which are the ones of a Maximum Likelihood (ML) reconstruction, as in \cite{Voyez-a-14}. 
If the regularization hyper-parameter $\mu$ is set satisfactorily, as done in practice by the method described in the Appendix, then Eq.~(\ref{eq-error-covariance}) yields  conservative and reasonable error bars. These error bars are displayed in the paper and are called MAP error bars.
\revision{Assessing the accuracy of these error bars is still a subject of ongoing research~\cite{efron_frequentist_2015} which is beyond the scope of the present paper.}

\section{Scindar data checks and an unsupervised $C_n^2$ profile reconstruction} \label{data-check}
This section presents preliminary data checks and first examples of reconstruction of $C_n^2$ profiles from Scindar data. In Subsection \ref{datacheck} \revision{data (slope and scintillation) computation is detailed, and consistency between data statistics and turbulence statistical models, described in Subsection \ref{formalism} and assumed for the reconstruction, is checked.}
Subsection~\ref{pseudo} describes and illustrates the structure of auto- and cross-correlations of slopes and scintillation so as to confirm the data quality.  
Subsection~\ref{hyper} illustrates our $C_n^2$ profile reconstruction strategy including the unsupervised choice of the regularization hyper-parameter. 

\subsection{Shack-Hartmann data quality checks}\label{datacheck}

\revision{Time series of slopes and scintillation indices are computed, for each frame, on two moving windows centered on the SHWFS images of each of the two sources. These moving windows therefore allow to separate the two source contributions and perform windowing to extract relevant pixels so as to improve the effective signal-to-noise ratio. Wavefront phase and scintillation statistics are then deduced from these data. We now check their consistency with the turbulence model assumptions of the CO-SLIDAR method.}

\revision{The verification of the Kolomorov model performed in this section is based on a least square reconstruction, from slope data, of the Zernike coefficients corresponding to $20$ Zernike modes (piston excluded), which corresponds to radial orders $n$ from $1$ to $6$. Time series of such coefficients are reconstructed from 1 minute slope records. We then compute the modal variances of the Zernike coefficients before computing the average variance over all modes of a given radial order.}

\begin{figure}[!htb]
    \centering
    	\begin{center}
		\begin{tabular}{c}
		\includegraphics[trim = 2.9cm 1.4cm 3.cm 1.5cm, clip,width=1\textwidth]{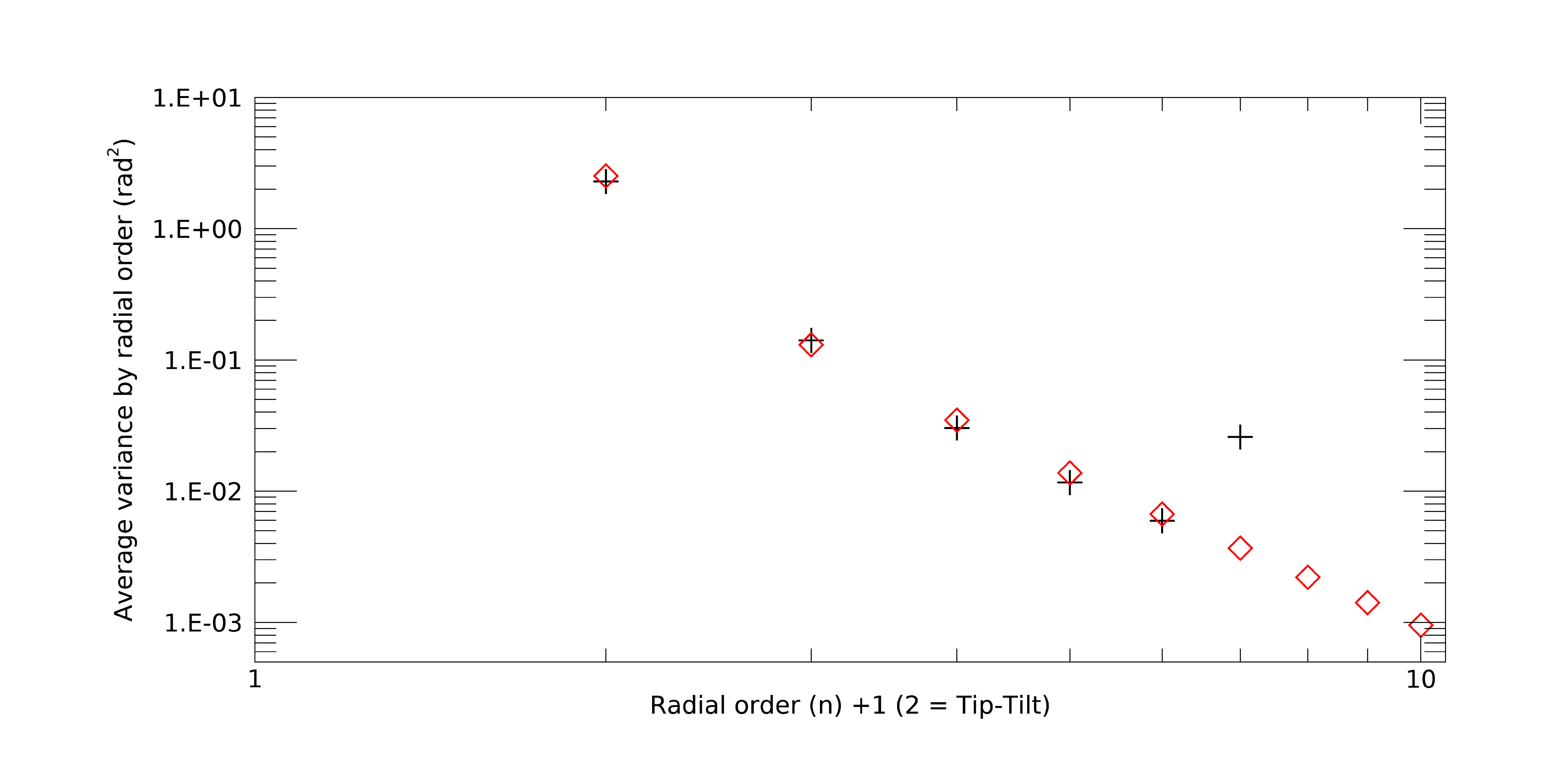} \\
	  
		(a) \\
		\includegraphics[trim = 1.5cm 0.4cm 2.8cm 1.cm, clip,width=1\textwidth]{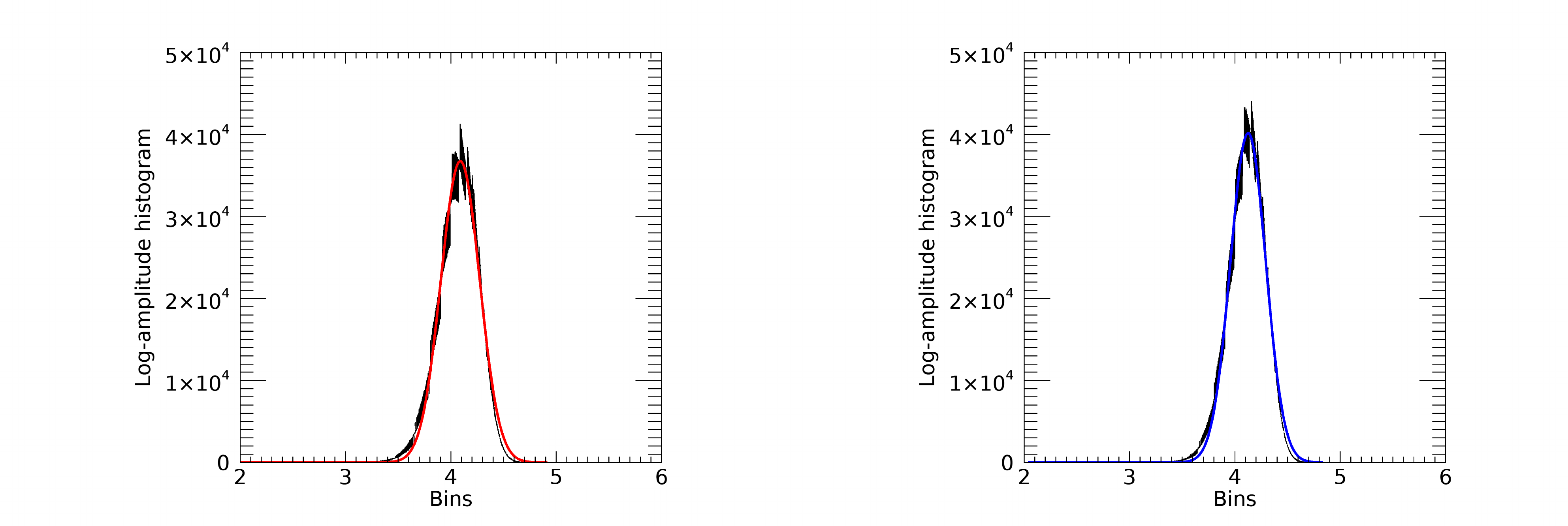} \\
	    (b) 
	        \end{tabular}
	\end{center}
 \caption{(a) Empirical Zernike variances averaged per radial order, computed over 1 minute of Shack-Hartmann slope data at 14:15 (black cross). Kolmogorov variance model (red diamonds) parametrized with the Fried parameter $r_0$. (b) Log-amplitude histogram of the top (left) and bottom (right) sources for the sequence starting at 14:15 over a 100-minute duration. Gaussian fit in colour.}   
 \label{fig:f5}
\end{figure}

\revision{Figure \ref{fig:f5}a compares this 
averaged variance as a function of $n+1$ (black crosses), and the Kolmogorov model variance (red diamonds). For radial orders $1$ to $5$, we observe a very good match with the Kolmogorov turbulence model. The fact that the match is good also for Tip and Tilt ($n=1$) confirms that there is no need to consider a finite outer scale in the model.}

As expected for Kolmogorov turbulence, the variance, averaged by radial order, decreases asymptotically as $(n+1)^{-11/3}$. \revision{The aliasing effect however corrupts  the reconstructed Zernike variance of radial order $n=6$. We recall that aliasing is induced by the finite spatial sampling of the wavefront by the SHWFS subapertures.}

The Kolmogorov variance model (red diamonds on the Fig~\ref{fig:f5}a) is parametrized with the Fried parameter $r_0$. More precisely, the Noll theory \cite{Noll1976} gives a relationship between $D/r_0$, where D is the telescope diameter, and the variance $\sigma_{n}^2$ of the  coefficient of any Zernike polynomial of radial order $n$, which can be written in the following way:
\begin{equation}
	\left(\frac{D}{r_0}\right)^{5/3}=\frac{\sigma_{n}^2}{(n+1)}\frac{\left[\Gamma\left(\frac{17}{6}\right)\right]^2\Gamma\left(n+\frac{23}{6}\right)}{\revision{2.246}\,\Gamma\left(n-\frac{5}{6}\right)},	\label{eq:DsurR0}
\end{equation}
where $\Gamma$ is the gamma function. 
The variance of each radial order yields, through Eq.~(\ref{eq:DsurR0}), an estimate of $({D}/{r_0})^{5/3}$. The average of the latter estimates for all radial orders between $n=2$ and $n=5$ is our final estimate of $({D}/{r_0})^{5/3}$. We shall plot this direct estimation of $r_0$ from SHWFS data in Subsection \ref{chroniques} for the two sequences studied.

The distribution of the log-amplitude $\chi$  is computed by applying the logarithmic function, on long time series, of the source intensity. Then we plot in Fig.~\ref{fig:f5}b the log-amplitude histogram for each source and fit a Gaussian probability distribution. We can observe that the Gaussian fits are very close to the log-amplitude distribution. For these records, the subaperture averaged scintillation indices $\sigma_i^2$  estimated on both sources are $0.13$ and $0.11$. They are far smaller than $1.2$, generally considered as the Rytov limit; we thus confirm the hypothesis of the weak perturbation regime. We will confirm that this regime is valid  for the two sequences studied in Subsection~\ref{chroniques}, where we plot the scintillation index averaged on the two sources, obtained directly from SHWFS data.

These verifications (Kolmogorov turbulence and weak perturbation regime) confirm the data consistency with the Scindar measurement model.

\subsection{Slopes and scintillation correlations quality check}\label{pseudo}

\begin{figure}[!htb]
    \centering
    	\begin{center}
		\begin{tabular}{m{1.8cm}m{1.8cm}m{1.8cm}m{1.8cm}m{1.8cm}m{1.8cm}}
		
		\multicolumn{6}{l}{	\includegraphics[width=12cm]{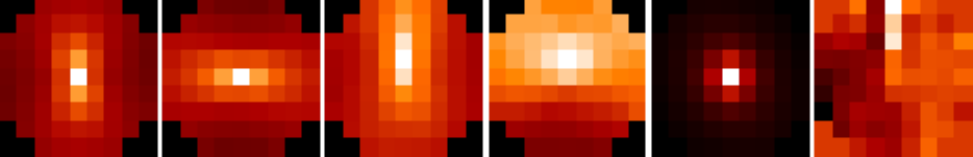}} \\
		\,\,\,\,\,\,\,\,\,\,\,\,\,(a) & 	\,\,\,\,\,\,\,\,\,(b) & 	\,\,\,\,\,\,\,(c) & \,\,\,\,(d) & (e) & (f)\\
		\end{tabular}
	\end{center}
    \caption{Correlation maps from scintillation and slopes measured with Scindar at 14:15 over a 3-minute recording (25560 frames). From left to right: (a) x-slopes auto-correlation of one source, (b) y-slopes auto-correlation of one source, (c)  x-slopes cross-correlation of \revision{two sources}, (d) y-slopes cross-correlation of two sources, (e) scintillation auto-correlation of one source,
    (f) scintillation cross-correlation of two sources.
    }
	\label{fig:f6}
\end{figure}

Before estimating the $C_n^2$ profiles, we plot the correlation maps of slopes and scintillation to describe their structure and compare it to the expected one. An example of correlation maps is presented on the Fig.~\ref{fig:f6}. These maps show the correlation averaged over all pairs of subapertures with a given separation (from Eqs.~(\ref{eq:16}-\ref{eq:18})). 
We actually show the average of the two auto-correlation maps, obtained for each of the source.
The auto-correlation map of x (respectively y) slopes on Figs.~\ref{fig:f6}a(-b) presents stronger correlations along y (respectively x) similar to the ones obtained in simulation. The cross-correlation map of x-slopes on Fig.~\ref{fig:f6}c shows - as expected - significant correlations in the direction of the source alignment (y): the peak of correlation associated to a given turbulent slice at a distance $z$ is shifted along the vertical axis by $\gamma(z) z$ (Eq.~(\ref{translation})).
The auto-correlation of scintillation is presented on Figure \ref{fig:f6}e and it appears narrower than that of slopes. Its width is related to the scintillation characteristic size $\sqrt{\lambda z  L/(L-z)}$ for a single slice as already mentioned on the Figure \ref{fig:poid}.  
As for the slopes cross-correlation maps (Fig.~\ref{fig:f6}c), the scintillation cross-correlation map (Fig.~\ref{fig:f6}f) shows, as expected, significant correlations in the direction of the source alignment $y$, despite a higher convergence noise. All maps of Fig.~\ref{fig:f6} depend on the $C_n^2$ profile. The cross-correlation maps is very informative for the reconstruction, mainly in the triangulation range $z=0$ to $z_{max}$. Scintillation auto-correlation also contain information on the slice position without range limitation (see Sect.~\ref{formalism}).

Let us explain briefly how the slopes correlation maps (Figs.~\ref{fig:f6}a-d) are corrected of the bias induced by instrumental vibrations. The SHWFS in the Scindar has an infrared camera with a cryo-cooler, the engine of which produces a narrow vibration peak centered on $50$ Hz. We reasonably assume that it affects evenly the slopes of all subapertures. We identify the energy of the vibration peak on the temporal power spectral density of the slopes. Its removal is then performed by subtracting the vibration variance (one tenth of $rad^2$ at most) on each point of the four slope correlation maps. 
The estimation of the vibration energy is done using the same amount of records as the correlation maps.

The quality of our experimental data being confirmed by the data checks presented in Sects.~\ref{datacheck} and \ref{pseudo}, we can now proceed with the reconstruction of the $C_n^2$ profiles. 

\subsection{Reconstruction of $C_n^2$ profiles}\label{hyper}
In this section, the GCV-based unsupervised adjustment of the hyper-parameter presented in Sect.~\ref{coslidar} is applied to estimate the $C_n^2$ profiles from the Scindar data. Every 3 minutes, the GCV function $V(\mu)$ (see Eq.(\ref{GCV}) of the Appendix) is calculated from a set of estimated $C_n^2$ profiles obtained with the corresponding set of values of the hyperparameter $\mu$. The reader can find in the Appendix a brief presentation of the GCV method and its validation for the problem at hand. 

\begin{figure} [!htb]
	\begin{center}
		\begin{tabular}{c}
			\includegraphics[trim = 4cm 1.7cm 2.7cm 2.3cm, clip,width=1\textwidth]{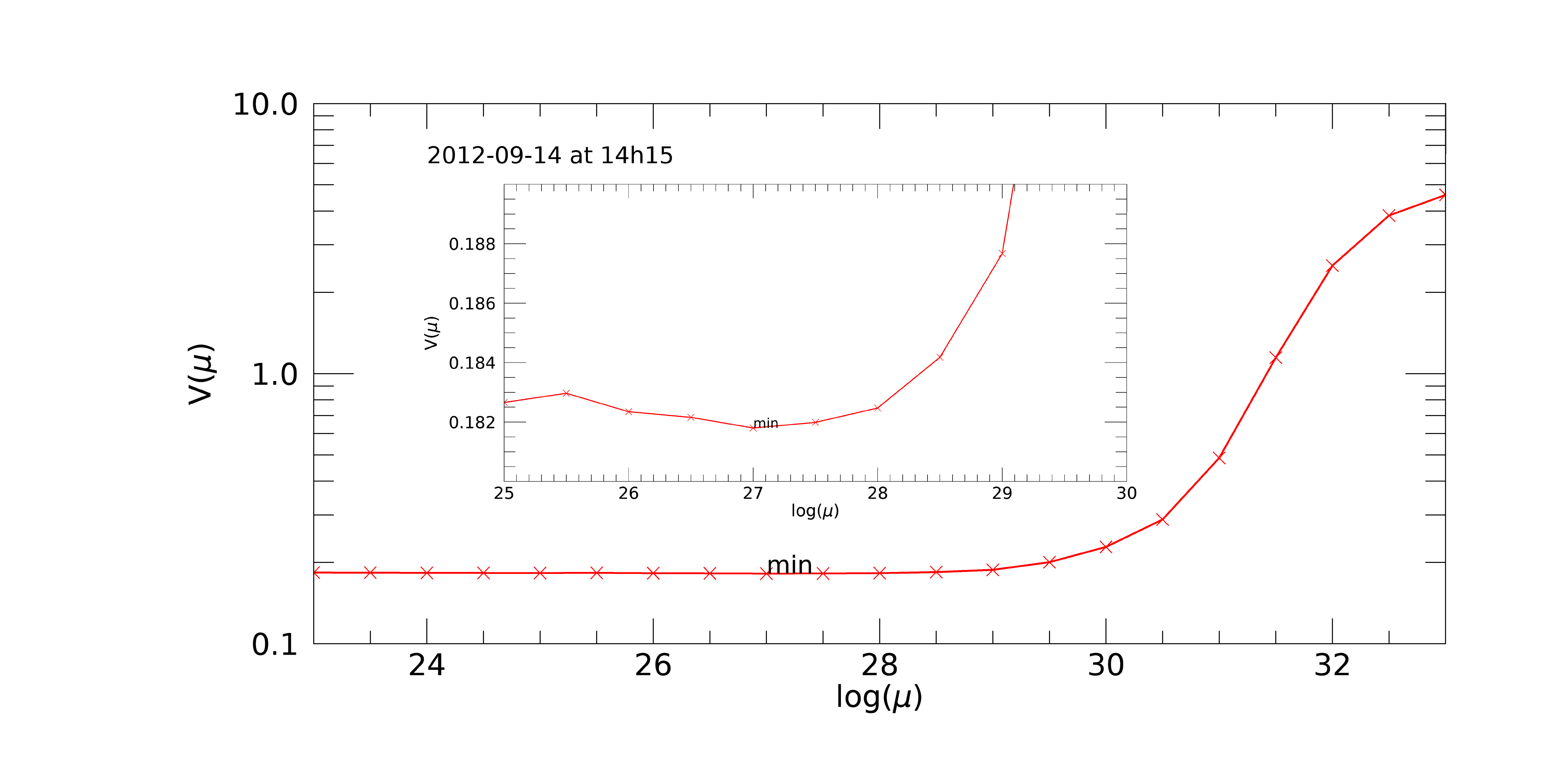} \\
			(a) \\
			\includegraphics[trim = 4cm 1.7cm 2.7cm 2.3cm, clip,width=1\textwidth]{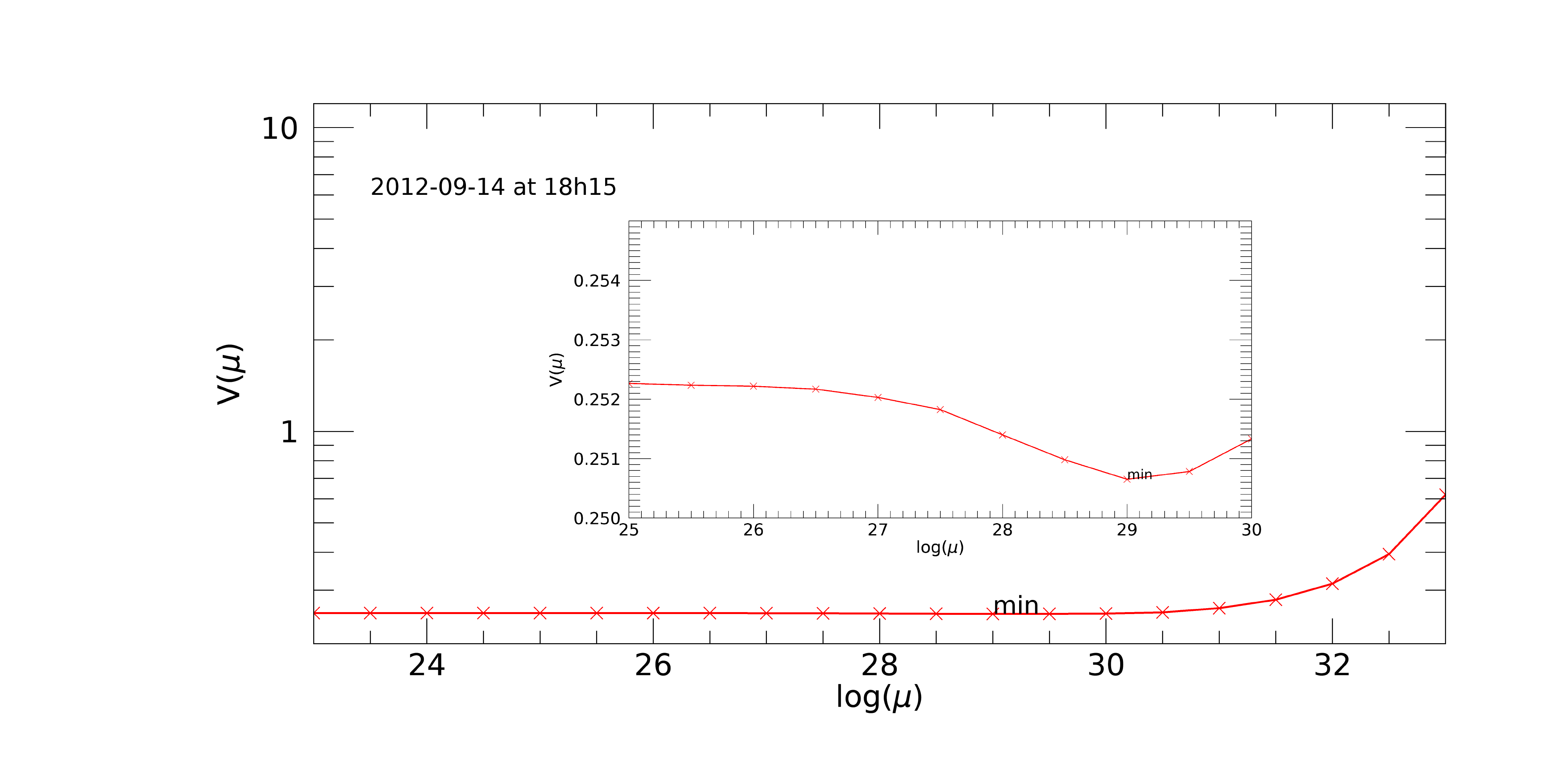} \\
			(b) 
		\end{tabular}
	\end{center}
	\caption {GCV function versus the regularization parameter $\log(\mu)$ for two  datasets (a) 14:15 and (b) 18:15. A zoom of the curve is inserted to better visualize the location of the minimum: (a) $\log(\mu)=27$ and (b) $\log(\mu)=29$.}
	\label{fig:TPRE}
\end{figure}

Figure \ref{fig:TPRE} presents two examples of GCV functions obtained respectively with the data recorded at 14:15 and 18:15. Their minimum values are respectively $\log(\mu)=27$ and $\log(\mu)=29$. As explained in Subsection \ref{regularization}, $\mu$ is the inverse of the \textit{a priori} variance of the $C_n^2$ values. The hyper-parameter $\mu$ obtained at 18:15 is a hundred times larger than that at 14:15, and consistent with the strong turbulence decline by a factor of ten of the $C_n^2$ average value along the line of \revision{sight} in the late afternoon.

\revision{Note that at 18:15, the turbulent parameters measured directly are $r_0=\,0.49$\,m and $\sigma_i^2=\,0.014$.}
\revision{As stated in Section~\ref{lannemezan}, 
all integrated parameters are given in the mid-infrared.}
The ones derived from the $C_n^2$ profile with the GCV-derived hyper-parameter are close: $r_0=0.41$ m and $\sigma_i^2=0.013$ and consistent with low turbulence conditions (the slight discrepancy in $r_0$ is discussed in Sect.~\ref{Cn2_profile}) and with the operating mid-infrared wavelength of the Scindar (Sect.~\ref{lannemezan}). 
On the start of the afternoon (14:15) the hyper-parameter $\log(\mu)$ is equal to $27.5 \pm 0.5$ during 25 batches of 3~minutes so 75~minutes. On the end of the afternoon (18:15) the hyper-parameter $\log(\mu)$ is equal to $29.5 \pm 0.5$ during 18 batches of 3~minutes so 54~minutes.
This consistency is an additional confirmation of the ability of GCV to produce reasonable hyper-parameter estimates.

\begin{figure}[!htb]
      	\centering
		\begin{tabular}{@{}c@{}}
		\includegraphics[trim = 2.5cm 1.4cm 3.0cm 1.7cm, clip,width=0.85\textwidth]{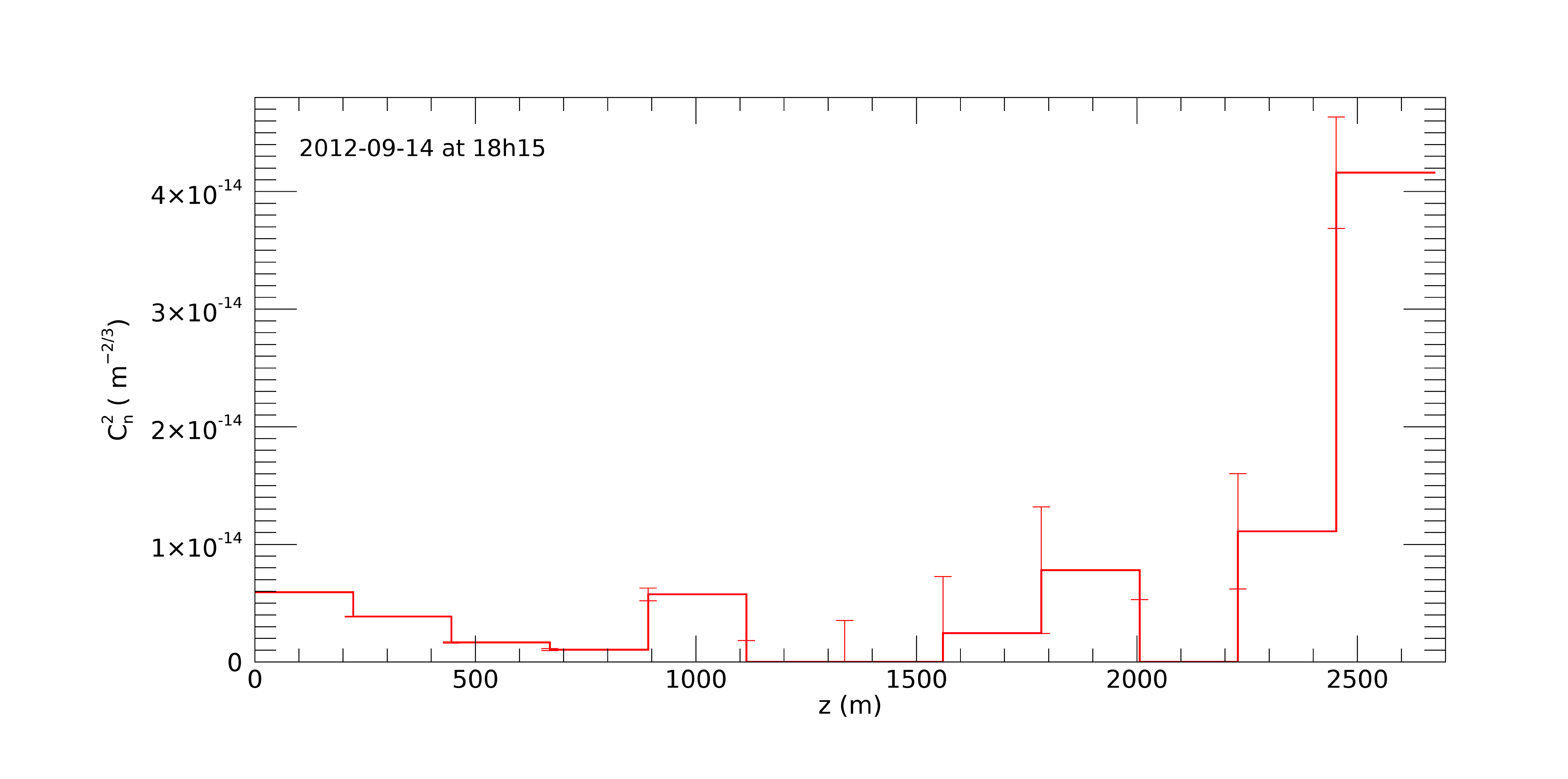}\\
		(a) \\
	    \includegraphics[trim = 2.5cm 1.4cm 3.0cm 1.7cm, clip,width=0.85\textwidth]{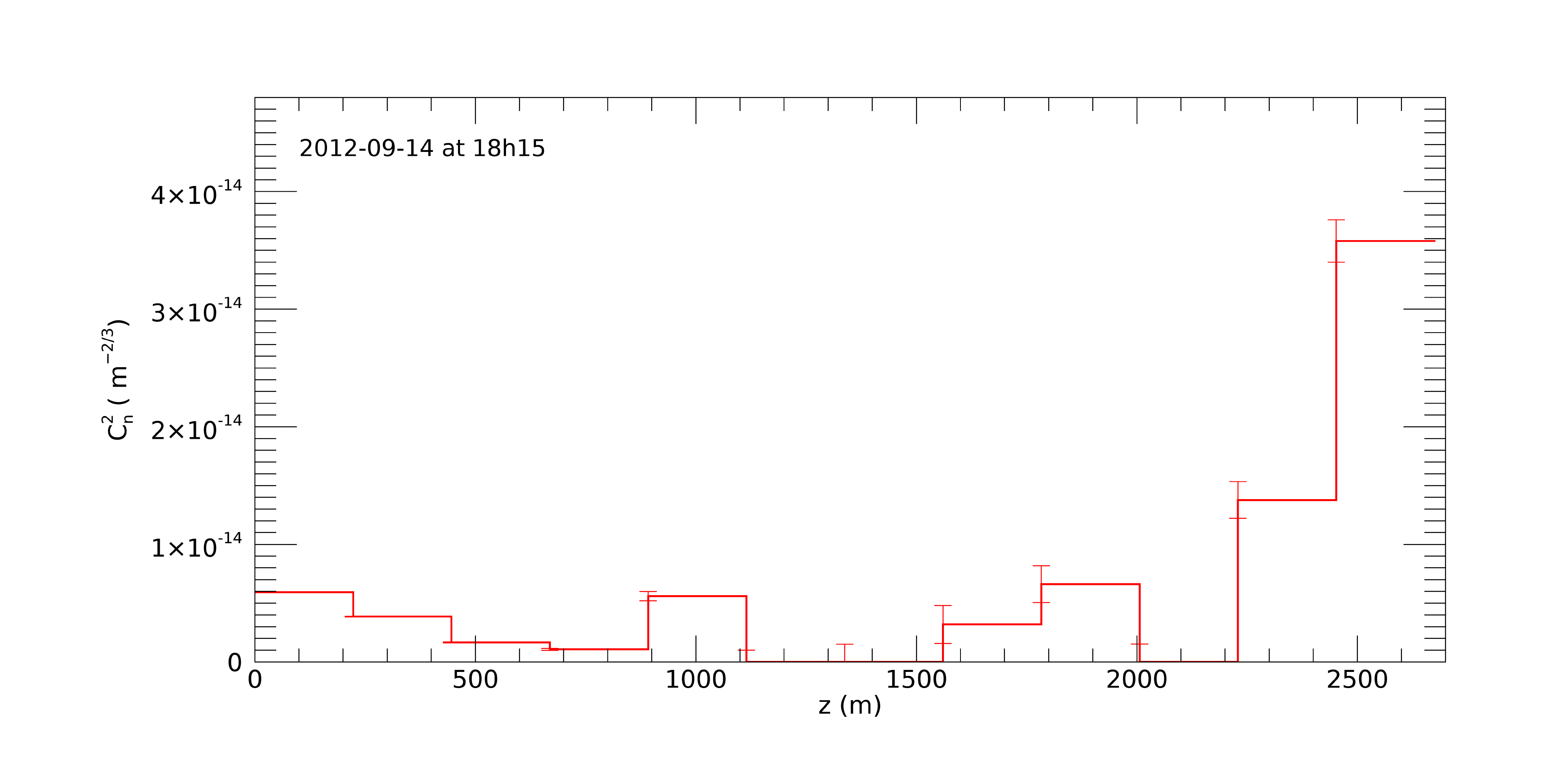}\\
	    (b) \\
	    \includegraphics[trim = 2.5cm 1.4cm 3.0cm 1.7cm, clip,width=0.85\textwidth]{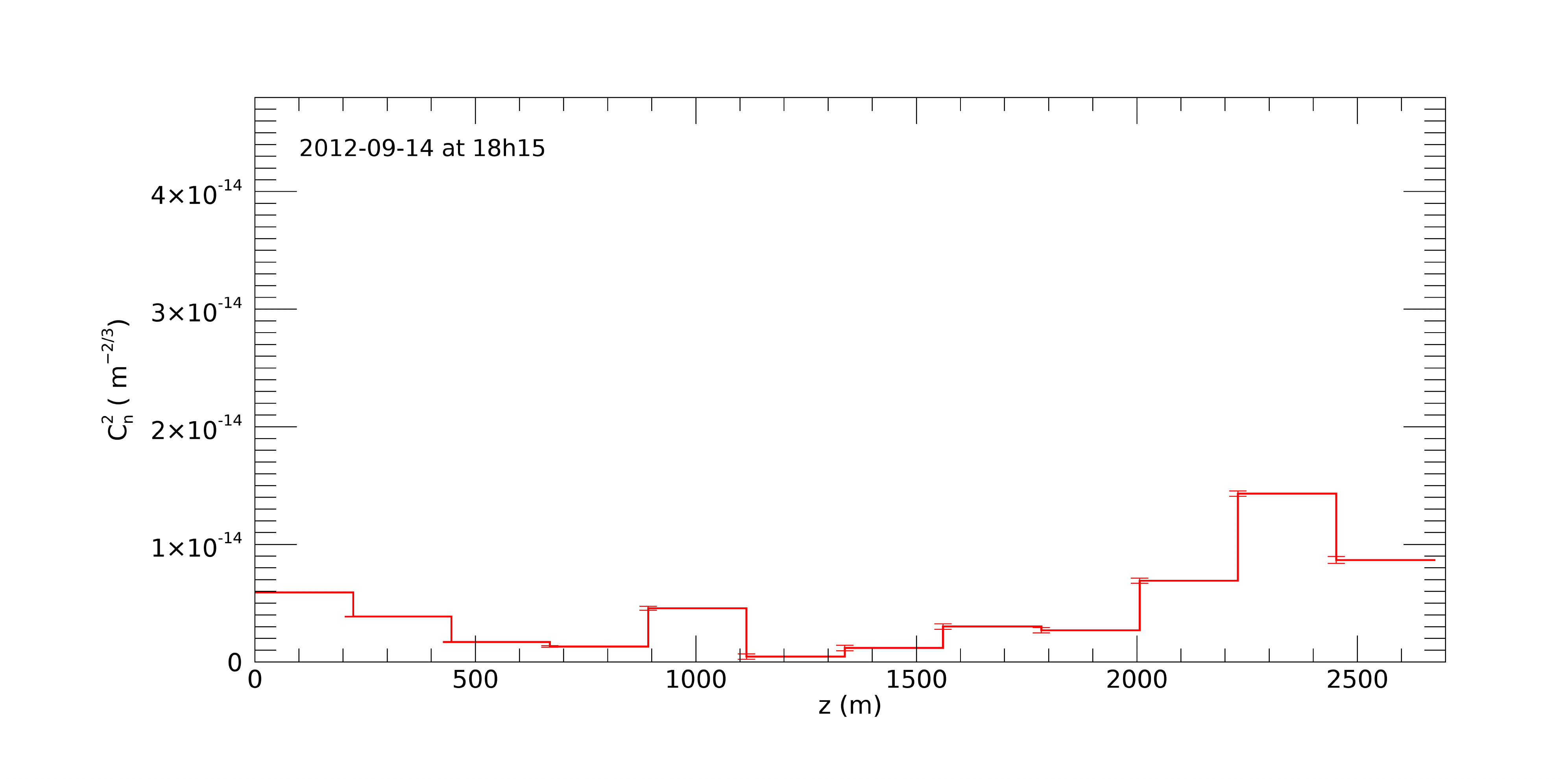}\\
	    (c)
	        \end{tabular}
    \caption{$C_n^2$ profiles reconstructed from the dataset recorded at 18:15 and for three hyper-parameters. (a) $\log(\mu) = 27$, (b) $\log(\mu) = 29$, (c) $\log(\mu) = 31$. The GCV-based reconstruction corresponds to (b).}
    \label{fig:f8}
\end{figure}

Figure \ref{fig:f8} shows three different $C_n^2$ profiles estimated from the 18:15 data set with different hyper-parameters adjustment.
Figure \ref{fig:f8}b presents the GCV-based regularization solution, Figs.~\ref{fig:f8}a and \ref{fig:f8}c present respectively a factor $100$ under/over regularization with respect to the GCV result. 
The plots of Fig.~\ref{fig:f8} confirm that, as is classical with regularized inverse problems, the sensitivity to the hyper-parameter is logarithmic (i.e., substantial changes of the reconstruction occur only for a change by a factor 10). 
Conversely, it demonstrates the importance of the hyper-parameter selection to better than a factor 10 and the interest of an unsupervised adjustment. 
The reconstructions and the associated error bars of Fig.~\ref{fig:f8} also suggest that, in this experiment, over-regularization is more harmful than under-regularization, as it reduces the reconstruction variance at the cost of a bias of the reconstruction towards zero, and this bias is quite strong close to the sources.

\clearpage

\section{$C_n^2$ profiling in heterogeneous rural landscape} \label{Cn2_profile}
This section presents the temporal evolution of the turbulence during the afternoon of 14/09/2012, as observed with the Scindar and the scintillometers 23~meters above the heterogeneous rural landscape.
We show in Sect.~\ref{chroniques} the temporal evolution of the turbulent parameters $r_0$, $\sigma_i^2$, and $\overline{C_n^2}$ weighted average value which are integrated quantities along the line of sight. 
We compare the latter with the scintillometers measurements and discuss their variability both in time and space. Then in Sect.~\ref{spatiotemp}, we present the $C_n^2$ profiles sampled every $223$ m together with MAP error bars.   

\subsection{Temporal evolution of turbulence parameters}\label{chroniques}
 The Fried parameter $r_0$ and the scintillation index $\sigma_i^2$ are key parameters that characterize atmospheric turbulence. 
These two parameters can be estimated in two ways: on the one hand, they can be estimated quite directly, from the data, independently of the reconstructed $C_n^2$ profile, as explained in subsection~\ref{datacheck}. On the other hand, they can be computed from the $C_n^2$ profile reconstructed with the approach described in Sect.~\ref{hyper}. Note that the scintillation index computed from the profile accounts for aperture and source averaging.

\begin{figure}[!htb]
    
    	\begin{center}
		\begin{tabular}{c}
		\includegraphics[trim = 3.5cm 1.4cm 3.1cm 2.0cm, clip,width=1\textwidth]{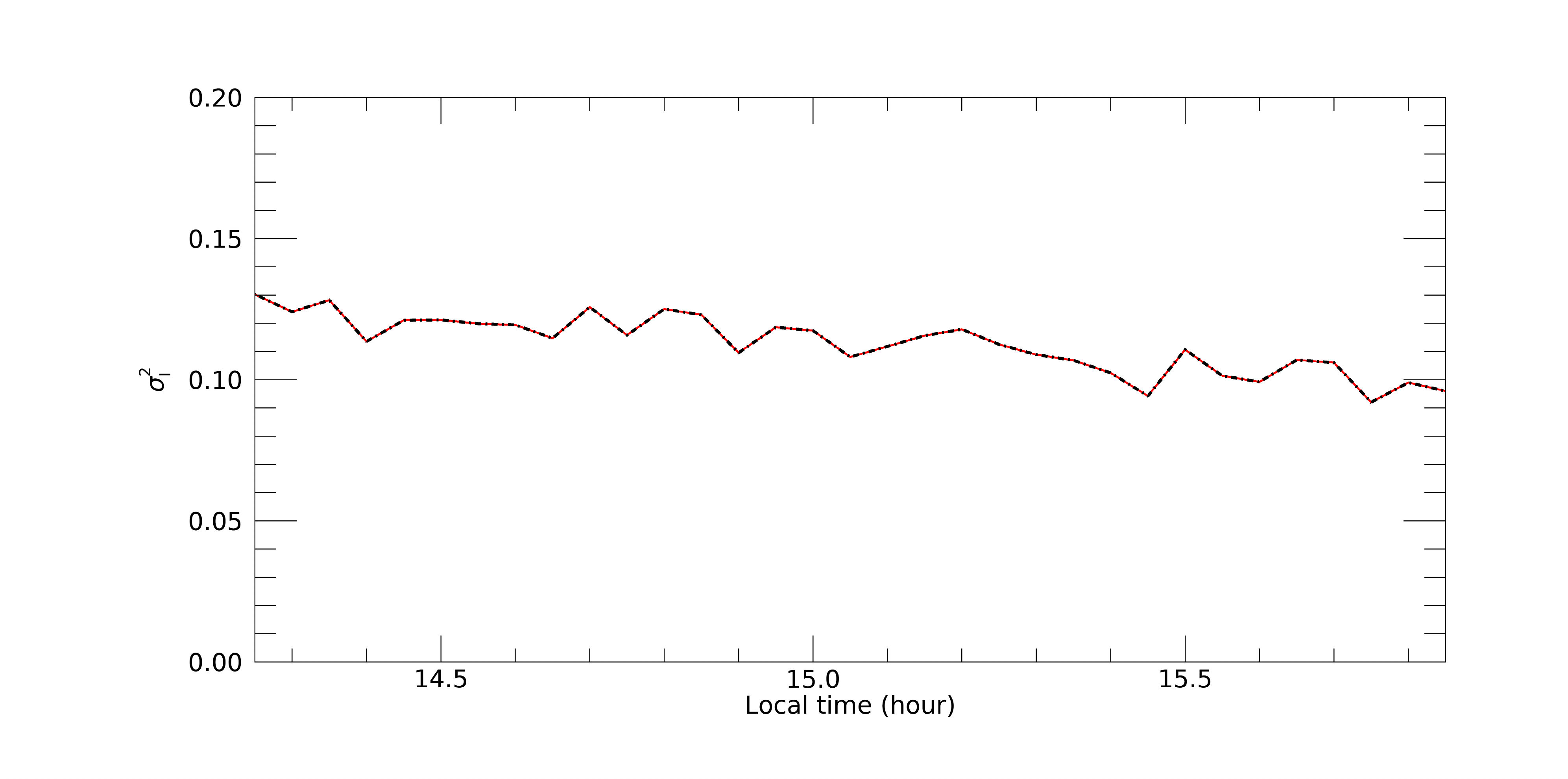}\\
		(a) \\
		\includegraphics[trim = 3.5cm 1.4cm 3.1cm 2.0cm, clip,width=1\textwidth]{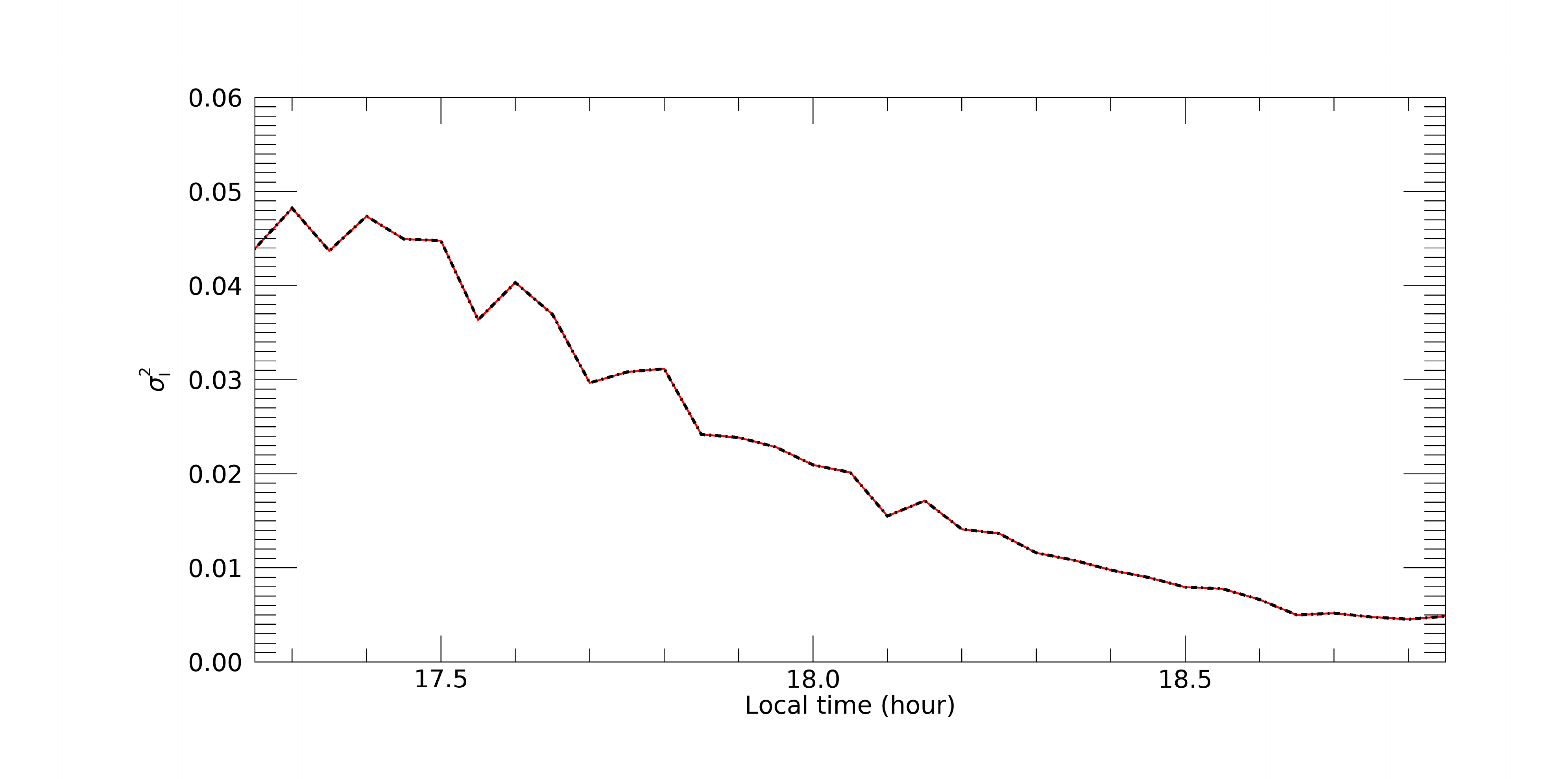}\\
	    (b) 
	        \end{tabular}
	\end{center}
 \caption{Temporal series of scintillation variance derived from the reconstructed profile with account of aperture and source averaging (red line) compared to the average scintillation variance of the two sources as measured directly on the SHWFS data (black dotted line): the two lines are almost superimposed. (a) at 14:15 during 100 minutes, (b) at 17:15 during 100 minutes.}   
 \label{fig:f9}
\end{figure}

\begin{figure}[!htb]
   
    	\begin{center}
		\begin{tabular}{c}
		\includegraphics[trim = 3.3cm 1.2cm 3.1cm 1.7cm, clip,width=1\textwidth]{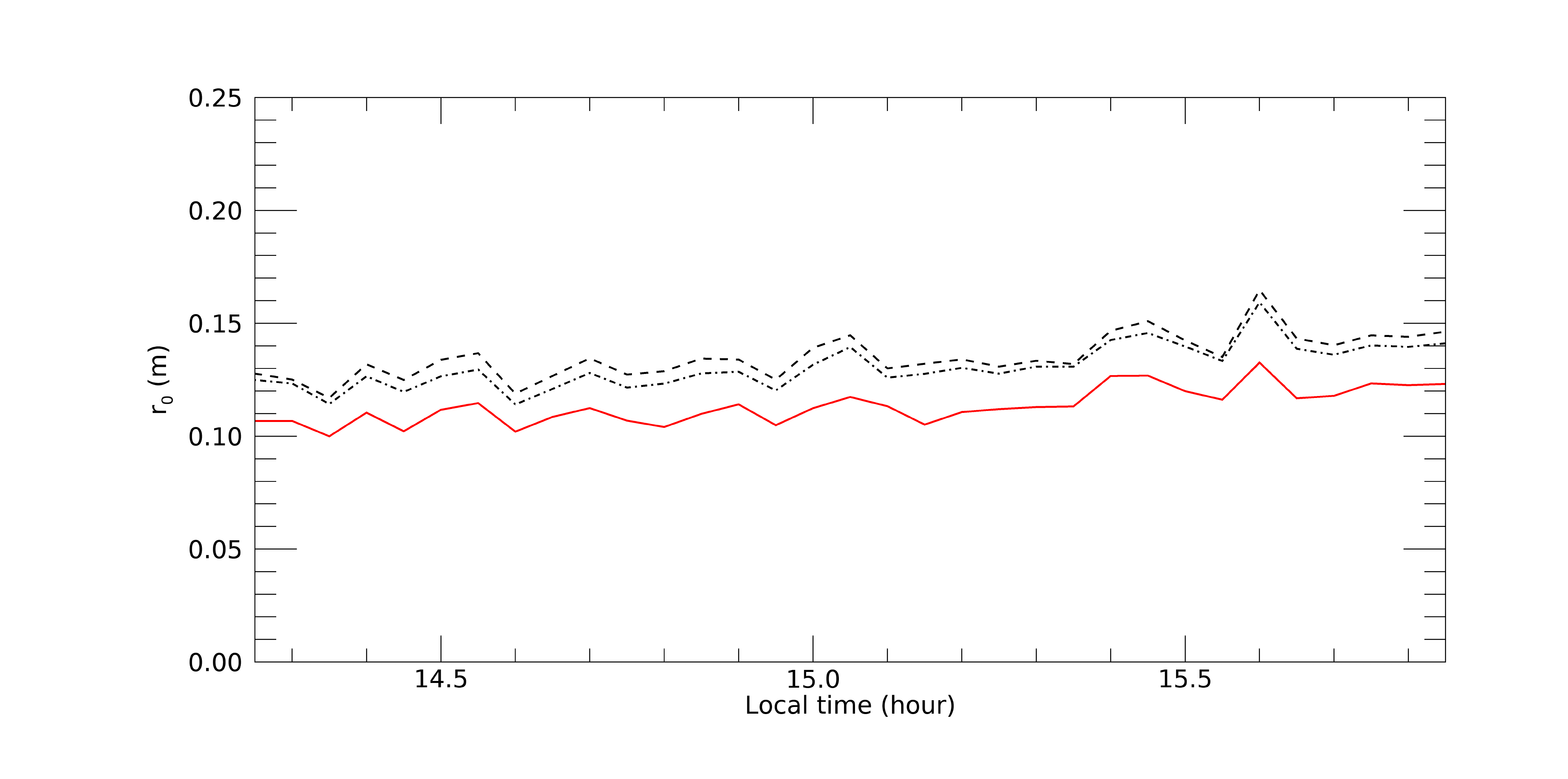}\\
	  
		(a) \\
		\includegraphics[trim = 3.3cm 1.2cm 3.1cm 1.7cm, clip,width=1\textwidth]{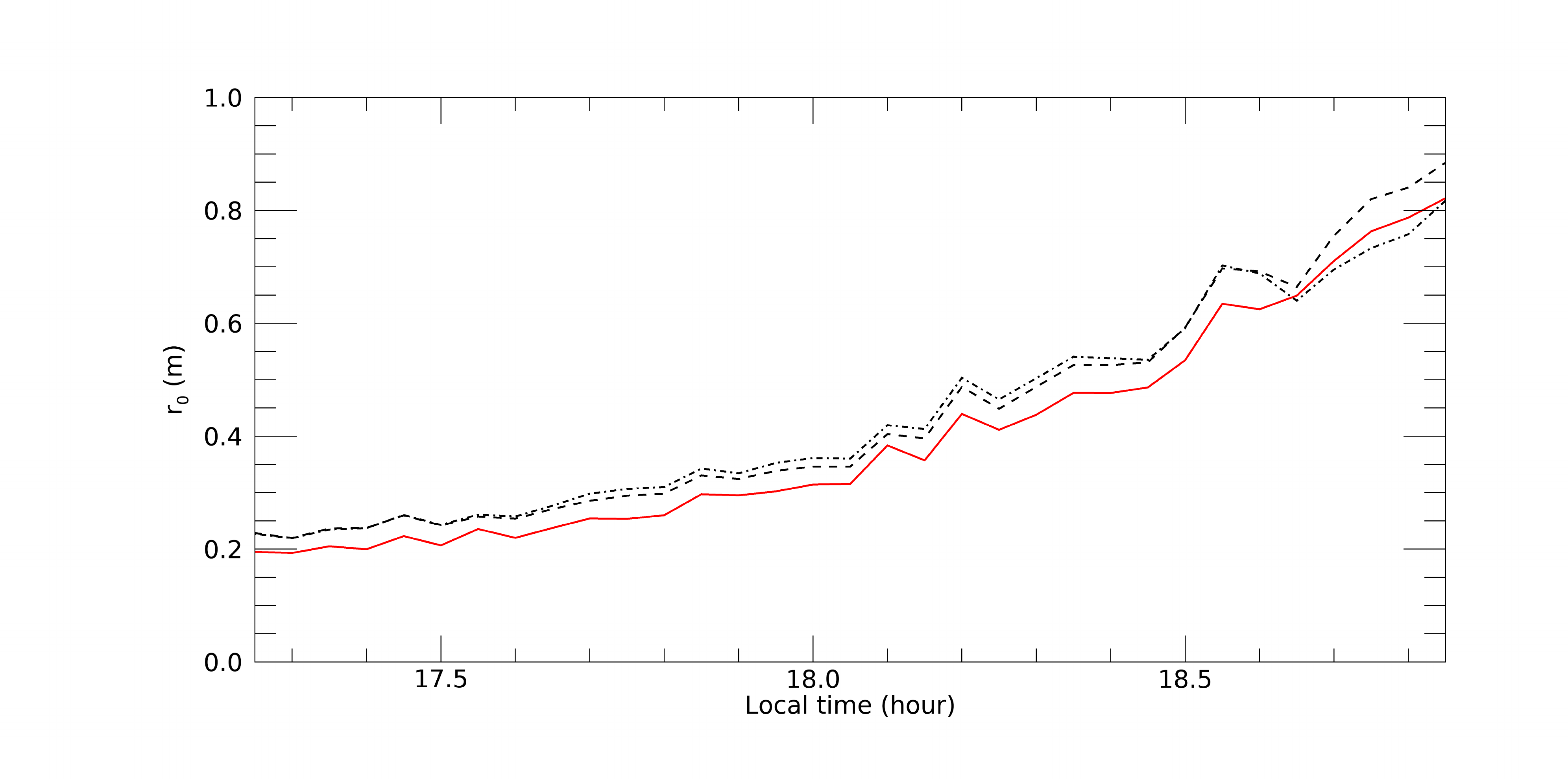}\\
	    (b) 
	        \end{tabular}
	\end{center}
 \caption{Temporal series of Fried parameter derived from the reconstructed profile (red line) compared to those estimated directly on SHWFS data for each source (two black dotted lines). (a) at 14:15 during 100 minutes, (b) at 17:15 during 100 minutes.}   
 \label{fig:f10}
\end{figure}

We plot on Figs.~\ref{fig:f9} and \ref{fig:f10} respectively the evolution of the variance $\sigma_i^2$ and of the Fried diameter $r_0$ during two sequences of 100 minutes roughly separated by one hour. The Figures show these turbulence parameters calculated from the $C_n^2$ profiles over-plotted with those estimated directly on the slopes and scintillation data. There is an excellent match  between the 2 estimations of $\sigma_i^2$ with very small relative errors ($RE$) for both sequences \revision{(for the first sequence, composed of 18 batches of 3 min, the temporal average of $RE$ is equal to $0.4$ with a standard deviation of $0.3$, and for the second sequence this average $RE$  is equal to $2.1$ with a standard deviation of $0.9$).}
However, for $r_0$, the value which is computed with the reconstructed profile is systematically smaller than that estimated from slopes. This may be due to the fact that SHWFS slopes are measured on extended sources in the presence of anisoplanatism: \revision{the phase perceived by the SHWFS (slopes and reconstructed wavefronts) are indeed apparent phases averaged over the angular extension of the sources leading to a weaker apparent turbulence and hence a larger apparent $r_0$. This bias is however not present in the profile reconstruction since the formalism of Subsection \ref{formalism} takes explicitly into account the source extension and the associated anisoplanatism effect.}

Turbulence features are different along both sequences on Figs.~\ref{fig:f9} and \ref{fig:f10}. At 14:15 and during 100 minutes, $r_0$ and $\sigma_i^2$ show very little variations, i.e., the turbulence can be considered stationary. In addition, we observe a stronger turbulence strength --- with lower $r_0$ values --- than for the late afternoon. At 17:15 and during 100 minutes, the turbulence strength declines, showing  a $r_0$ increase and a $\sigma_i^2$ decrease in time. 
These observations fit with the diurnal turbulence trends in clear sky and anti-cyclonic \revision{conditions \cite{stull_introduction_1988}.}

As already explained in subsection \ref{regularization}, the unsupervised tuning of the hyper-parameter gives the \textit{a priori} standard deviation of $C_n^2$ ($\sigma_{prior}$) which indicates an order of magnitude for the mean $C_n^2$. The latter mimics very well this trend: $\sigma_{prior}$ typically goes from  $10^{-13.5}$ during the early afternoon to values varying between $10^{-14}$ and $10^{-15}$ in the late afternoon.

\begin{figure} [!htb]
	\begin{center}
		\begin{tabular}{c}
			\includegraphics[trim = 2.6cm 1.1cm 2.3cm 2.2cm,clip,width=1\textwidth]{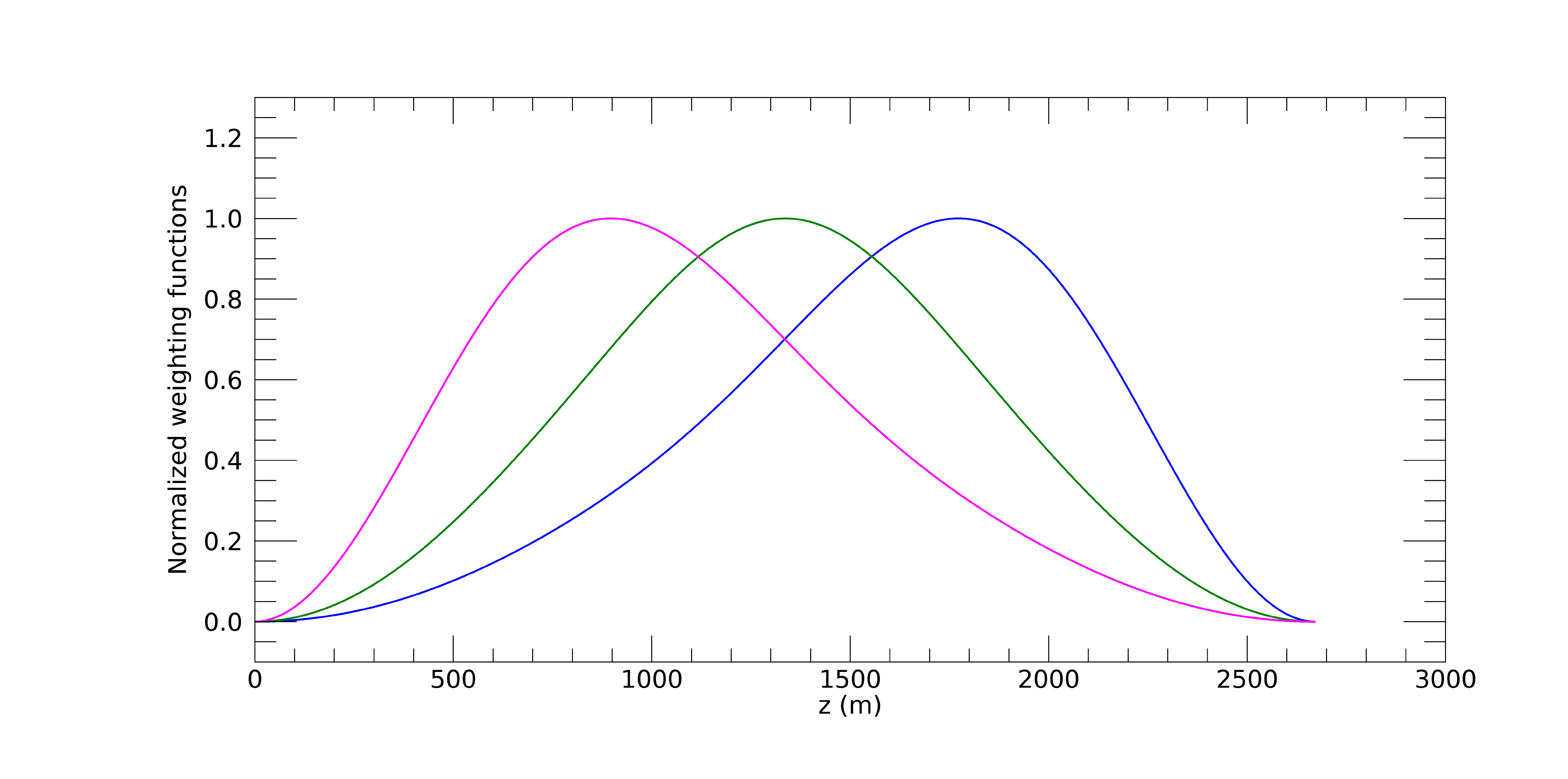}
		\end{tabular}
	\end{center}
	\caption {The normalized weighting functions of the three scintillometers as a function of $z$, i.e. the distance from the CRA. They are plotted with the pink, green and blue lines respectively for the A, B and C scintillometers.}
	\label{fig:fct_poid}
\end{figure}

We now turn to the scintillometer data. Scintillometers of the first generation have been designed to aggregate turbulent information over long distances. A scintillometer measures a $\overline{C_n^2}$ weighted average of the $C_n^2$ profile along the line of sight. The $\overline{C_n^2}$ value is deduced from the variance of the log-amplitude $\sigma_{\chi}^2$ recorded at a kHz sampling rate and averaged on 3-minute periods for consistency with the Scindar data. The scintillometer observation is $\left< \sigma_{\chi}^2 \right> = \alpha^{-1} \overline{C_n^2} D_t^{-7/3} L^3$, where $\alpha = 4.48$ for $D_t/D_r=1$ (scintillometer B) and $\alpha = 1.61$ for $D_t/D_r = 3$ (scintillometers A and C). 
We plot on Fig.~\ref{fig:fct_poid} the scintillometer weighting functions along $z$ - i.e. the distance to CRA. The peak location of each weighting function depends on the ratio of the scintillometer apertures size and determines $\alpha$. This should allow one to differentiate $\overline{C_n^2}$ observations representative of three different parts of the path.

\clearpage

\begin{figure} [!htb]
	\begin{center}
		\begin{tabular}{c}
			\includegraphics[trim = 2.6cm 1.5cm 2.cm 2.2cm, clip,width=1\textwidth]{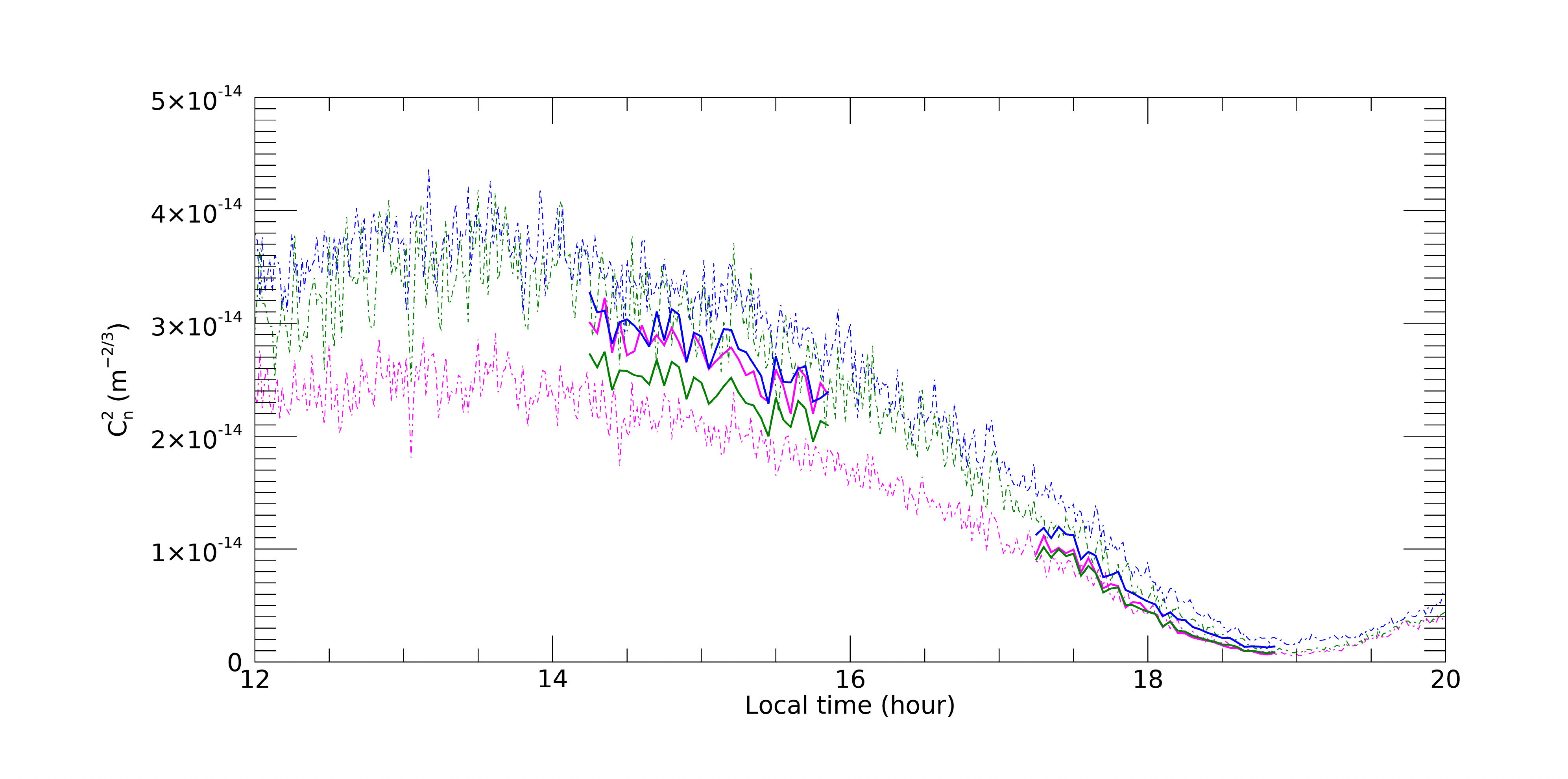}
		\end{tabular}
	\end{center}
	\caption {Temporal series of $\overline{C_n^2}$ weighted average values measured with the 3 scintillometers (dotted lines) compared to those deduced from Scindar $C_n^2$ profiles by applying the respective weighting functions (solid lines). The pink, green and blue lines correspond respectively to the A, B and C scintillometers.} 
	\label{fig:f12}
\end{figure} 
On the Fig.~\ref{fig:f12}, we show the $\overline{C_n^2}$ weighted average values measured by the scintillometers in dashed lines. The $\overline{C_n^2}$ weighted average emulated from the Scindar profiles by applying the respective scintillometer weighting functions (see Fig.~\ref{fig:fct_poid}) is plotted in continuous lines.  The values of $\overline{C_n^2}$ are of the same order of magnitude between the emulated ones and that from scintillometers. The relative differences between them are $30\%$ in the worst case. Finally, we can see that Scindar and scintillometer measurements both follow the diurnal cycle of the turbulence. Between 12h00 to 20h00, $\overline{C_n^2}$ values follow the solar radiation daily cycle with maximum values between 13h00 to 14h00. Then they decrease to reach a minimum around 19h00. In the evening negative sensible heat fluxes stratify the atmospheric boundary layer and $C_n^2$ is increasing again with turbulence associated  with shear.

We have presented the temporal evolution of spatially integrated turbulence parameters. Nevertheless, the Scindar --- as a new generation scintillometer --- has the key asset to provide discrete $C_n^2$ profiles along the line of sight at the landscape scale, that are presented in the next subsection so as to analyse  their spatio-temporal evolution.

\subsection{Spatial and temporal variability of $C_n^2$ along the path}\label{spatiotemp}
\begin{figure}[!htbp]
    	\centering
		\begin{tabular}{@{}c@{}}
		\includegraphics[trim = 2.cm 1.4cm 3.1cm 1.7cm, clip,width=0.63\textwidth]{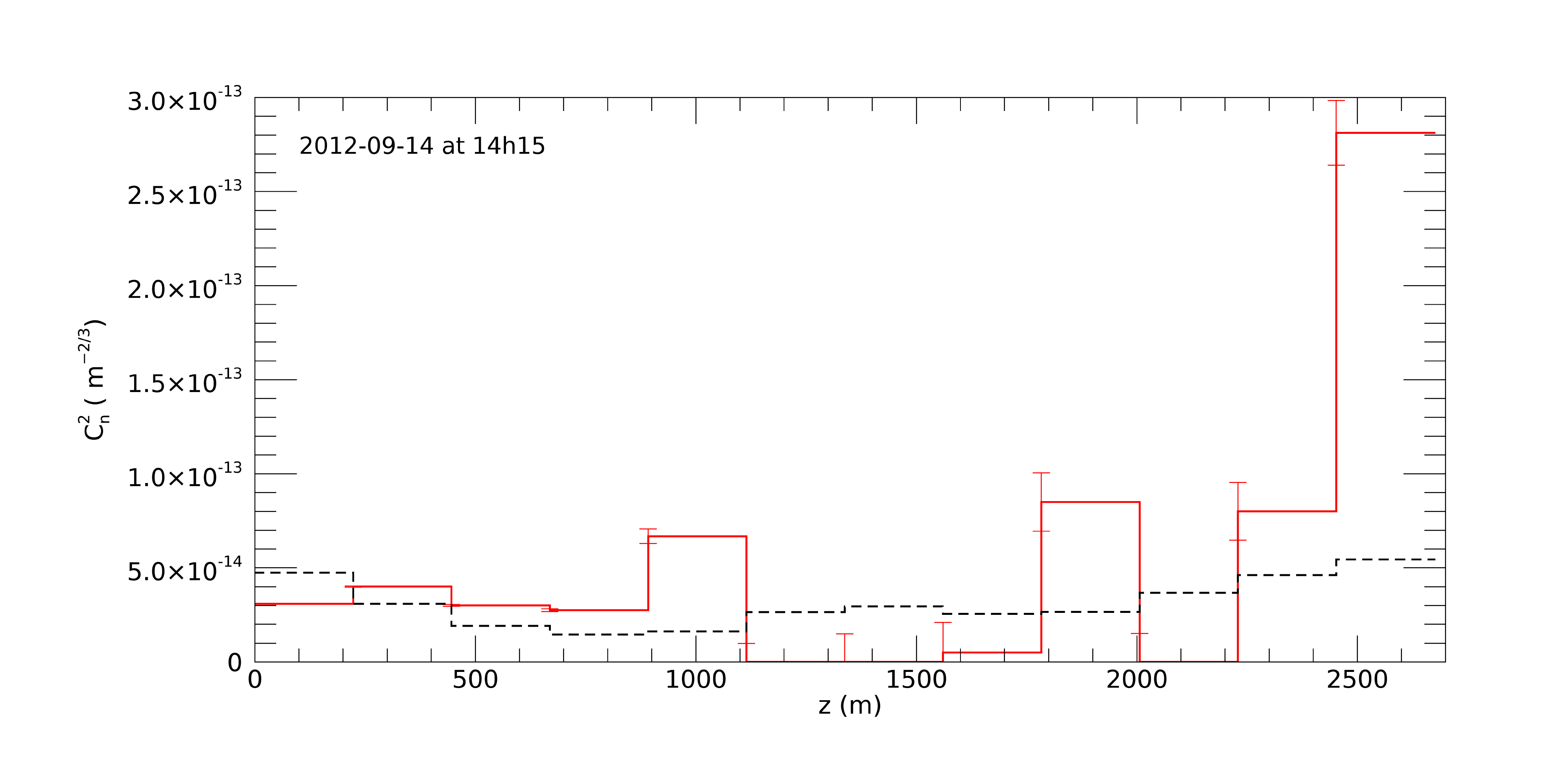} \\
	    	(a) \\
		    \includegraphics[trim = 2.cm 1.4cm 3.1cm 1.7cm, clip,width=0.63\textwidth]{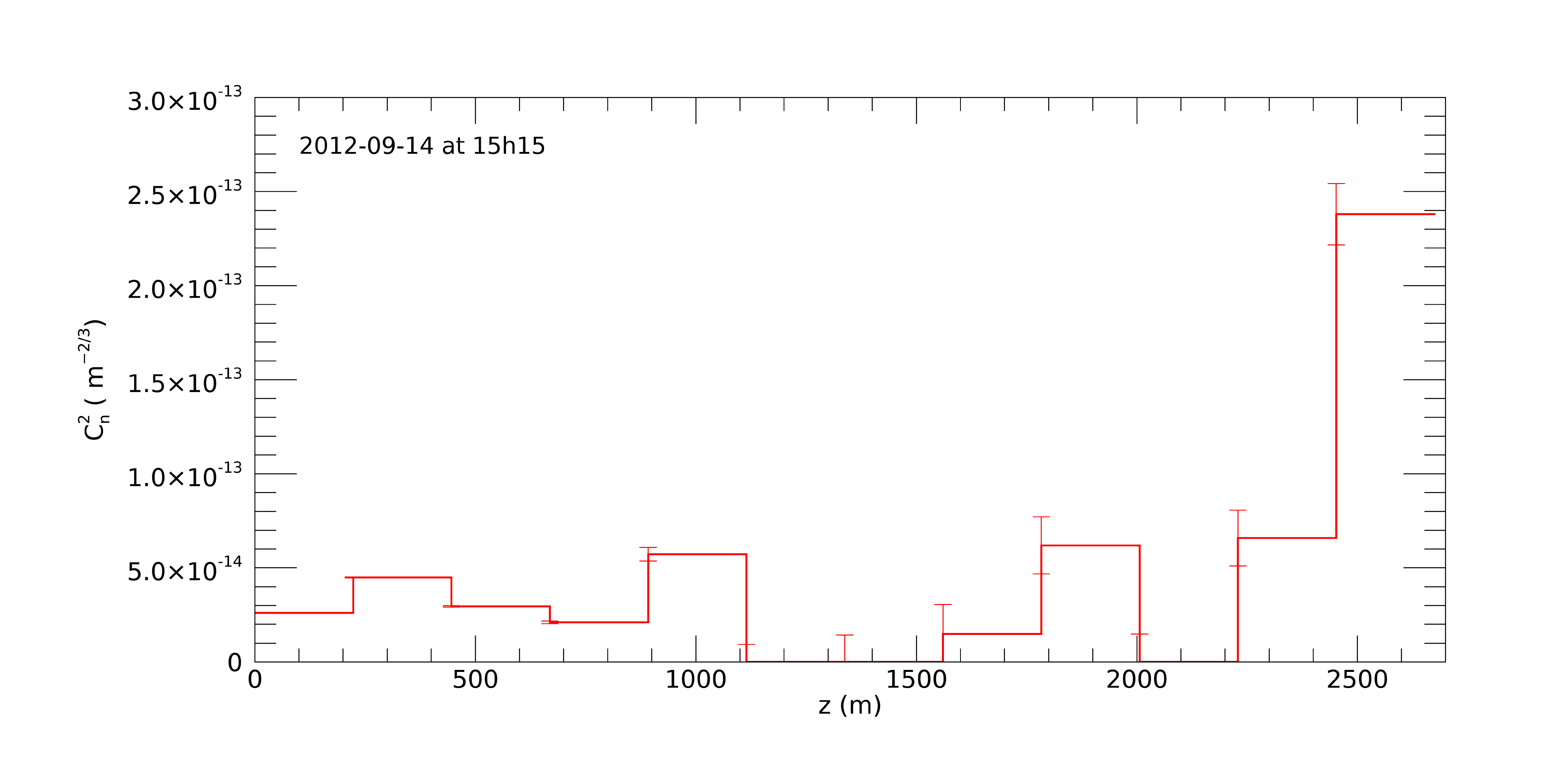} \\
	        (b)  \\
	        \includegraphics[trim = 2.5cm 1.4cm 3.1cm 1.7cm, clip,width=0.63\textwidth]{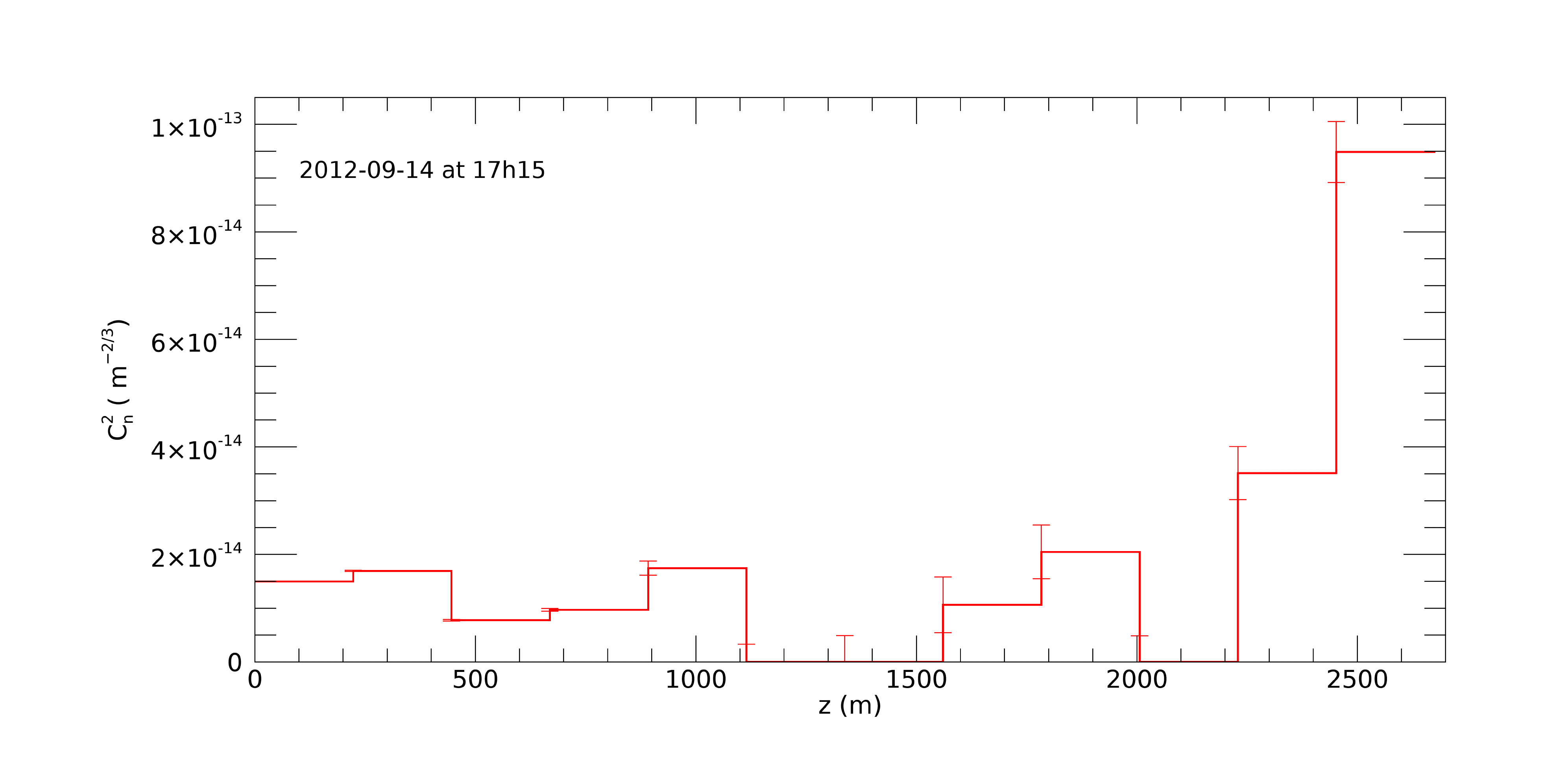} \\
	  		(c) \\
		\includegraphics[trim = 2.5cm 1.4cm 3.1cm 1.7cm, clip,width=0.63\textwidth]{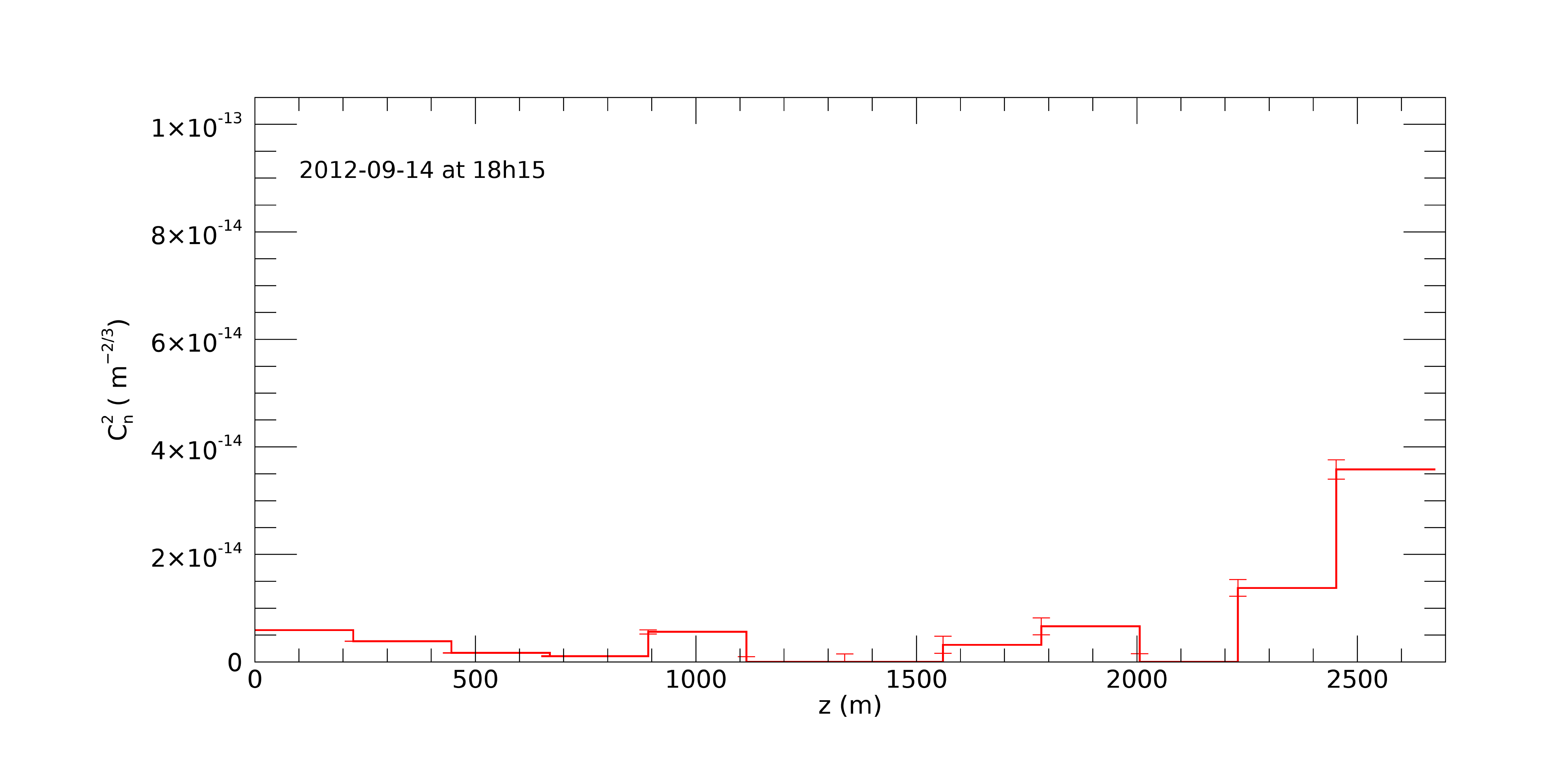} \\
	    (d) 
	        \end{tabular}
 \caption{Scindar $C_n^2$ profiles (red) estimated hourly in the afternoon of 14/09/2012. The black dashed line on graph (a) overplots a model of $C_n^2$ profile derived from the Monin Obukhov similarity theory in unstable diurnal conditions.}   
 \label{fig:f14}
\end{figure}

In this subsection, the variability of the profile is shown both in time and in space. Plots on the Fig.~\ref{fig:f14} display four examples of $C_n^2$ profiles obtained from 3 minutes of data with $3\times 8520$ SHWFS images at 14:15, 15:15, 17:15 and 18:15 (in local time). 
These $C_n^2(z)$ profiles with error bars are estimated with the unsupervised MAP criterion, the 12 slices of the profiles are separated by 223~m.
The error bars are relatively small compared with the spatial variability of the $C_n^2$ observed along the path. We can therefore try to link the $C_n^2$ values with the land use map.

The $C_n^2$ profiles show three parts. The first one contains small and quite uniform values located on the CRA side over an heterogeneous crop zone with trees ($0-1100$ m). 
The second one is located above a crop zone with hedgerows ($1100-2000$ m), it displays very low values of $C_n^2$ between $1100$ and $1500$ m.
This could be explained by a very low difference of temperature between the atmosphere and the ground \cite{Potvin2008}, and also by a small ground's rugosity length.
Finally, we observe higher $C_n^2$ values on the village side ($2000-2700$ m), which corresponds to mixed crops and houses, where higher temperature and rugosity are expected. 

On Fig.~\ref{fig:f14}a the black dashed line displays a model of $C_n^2$ profile derived from the Monin Obukhov similarity theory in unstable diurnal conditions. The $C_n^2$ depends on the height of the line of sight above ground $h$ according to the parametric law $C_n^2(h_0) \times h^{-4/3}$ with  $C_n^2(h_0=10$~m$)$ taken constant and equal to $10^{-14}$~m$^{-2/3}$. This value is inferred with the Fried parameter $r_0$ observed at 14:15. The height deduced $C_n^2$ profile is not consistent with the Scindar reconstructed profiles, since it lacks the consideration of land cover linked to the landscape rugosity.

\begin{figure} [!htbp]
	\begin{center}
			\includegraphics[width=1\textwidth]{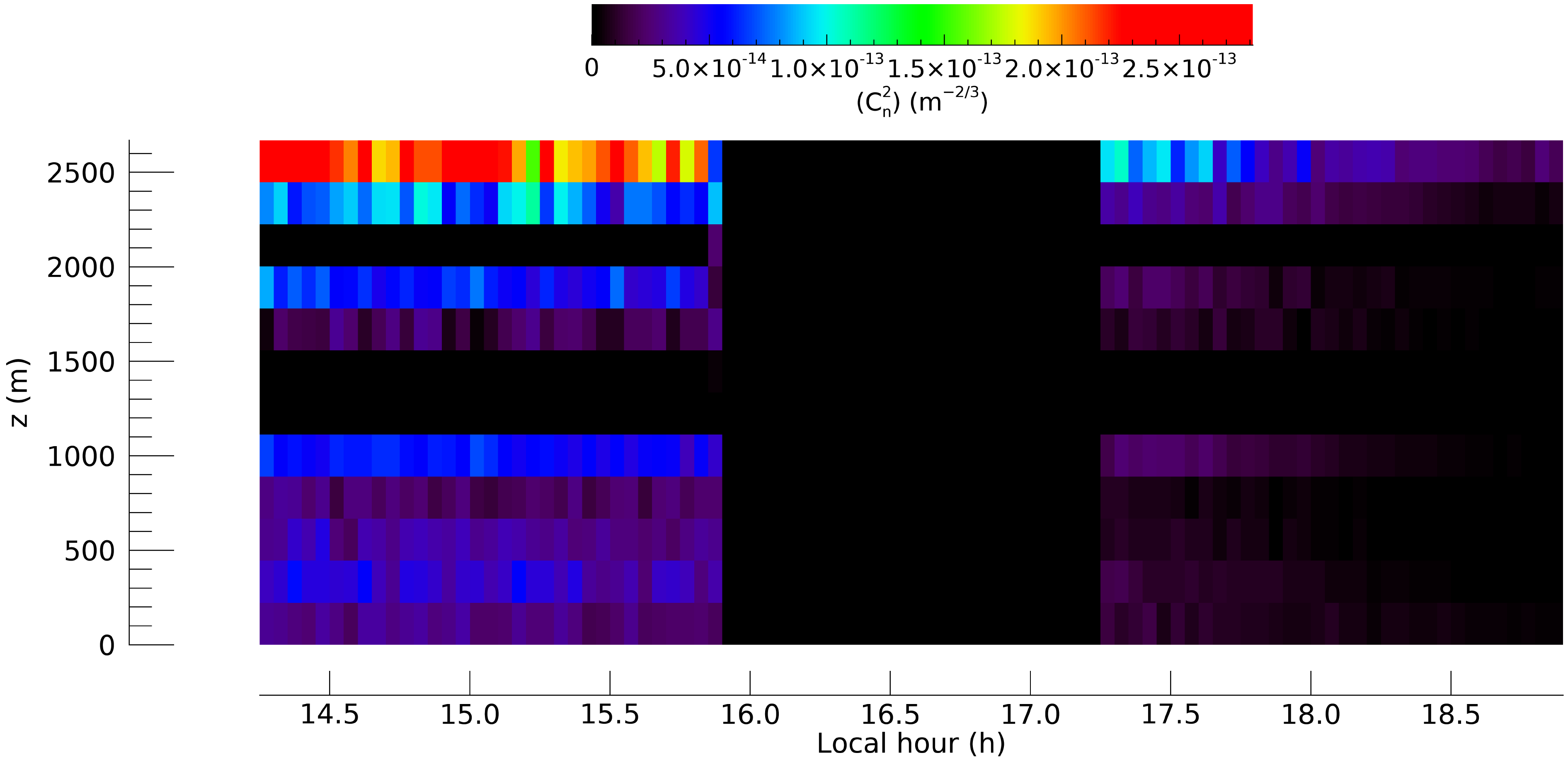}
			\end{center}
\caption {Temporal dynamics of $C_n^2$ along the optical path measured by Scindar during the afternoon of 14/09/2012.}
		\label{fig:f15}
\end{figure}

A spatio-temporal view of the main turbulent slices is shown on Fig. ~\ref{fig:f15}. The structure of the profile stays stable between 14:15 and 15:55. The turbulence strength declines in the late afternoon between 17:15 to 18:55. Along the afternoon, we observe the stable structure of the $C_n^2$ profiles with very low $C_n^2$ values systematically encountered in the $1100-1550$ range and in the $2000-2225$ range. These very low turbulence slices have a thickness equal to respectively the two or one discretization steps.  This indicates that the resolution of the Scindar is close to $223$\,m. Finally, the significant slices evolve with temperature led by the solar radiation and the terrain covers. The turbulence above the natural crop zones decline before that on the village as expected due to heat storage in the buildings.

\section{Conclusion} \label{conclusion}

Scintillometers of the first generation have been used for many applications over landscapes with smooth topography and/or heterogeneity. Unfortunately, terrain complexity introduces uncertainties and biases we cannot detect with a single aggregated measurement. For complex landscapes, there is a convergence towards the need for measurements with a resolution of a few hundred meters.

The CO-SLIDAR method, which exploits both phase slopes and scintillation, is a powerful tool for the metrology of near ground $C_n^2$ profiles, on the full line of sight between pupil and sources, with such a resolution. The Scindar, which is the associated transportable instrument, is based on a mid-IR Shack-Hartmann wavefront sensor \cite{Robert2012}, coupled to a 0.35-m telescope, which observes two cooperative sources. 
This paper presented the first comprehensive description of the CO-SLIDAR method in the context of near ground metrology. 
It included the presentation of the physical model, which accounts for the specific geometry: finite distance propagation hence spherical waves, proper aperture and source filtering. 
Furthermore, this paper developed a $C_n^2$ profile reconstruction strategy that is a Maximum \emph{A Posteriori} (MAP) solution with a white quadratic regularization. Two noteworthy assets of this strategy are the selection of the regularization hyper-parameter in an unsupervised way by using the GCV method, and the supply of error bars on the $C_n^2$ values. 
The reliability of the GCV method and the MAP error bars  have been validated on synthetic data.

The application to Scindar data acquired in a heterogeneous rural landscape during an experimental campaign demonstrated the ability to obtain a resolution of about $220$\,m over a $2.7$\,km line of sight with associated error bars. It confirmed the ability of the CO-SLIDAR technique to bypass the maximum distance related to triangulation.
The scintillometer-like measurements emulated from the reconstructed profiles are consistent with the measurements performed by the actual three commercial scintillometers.
The variability in space and time of the $C_n^2$ profiles was presented; it follows the diurnal cycle, and confirms the land cover influence.

As a perspective, we plan to work on the joint estimation of wind and $C_n^2$ profiles, which is of interest for the modeling of atmospheric physics at the ground/atmosphere interface, via the flux profiles entering the analytical footprint models~\cite{krapez_highly_2020} and the mesoscale models (Meso-NH)~\cite{robert_characterization_2019}.

\begin{backmatter}
\bmsection{Acknowledgments}
The authors wish to acknowledge assistance from Dr.\ J.~Voyez and training students Y. Tellier and A. Duong, technical work by F. Mendez and  F. Fleury, fruitful discussions with M.-T. Velluet, V. Michau and N. Védrenne. 
We express our gratitude to A.~Brut, F.~Lohou and S.~Derrien for their dedicated assistance during the experimental campaign.

\bmsection{Funding}
This study was made possible thanks to the financial support from ONERA (ECLAIR project).

\bmsection{Disclosures}
The authors declare no conflicts of interest.

\bmsection{Data availability}
Data underlying the results presented in this paper are not publicly available at this time but may be obtained from the authors upon reasonable request.

\end{backmatter}

\clearpage
\appendix 

\section{Appendix: Evaluation of the Generalized Cross Validation method and of the MAP error bars}

\subsection{Objectives}

The first aim of this appendix is to present and validate, on synthetic data, the method used to adjust the regularization hyper-parameter $\mu$ of our MAP $C_n^2$ profile reconstruction.
The chosen method is the Generalized Cross Validation (GCV), \revision{described along with other possible choices in~\cite[Sec.~2.3]{Demoment-l-08a}}. We shall show in the following that it allows an unsupervised and satisfactory adjustment of the regularization hyper-parameter. The second aim is to assess the reliability of the MAP error bars associated to the reconstructed profile.

\subsection{Direct problem and reduced data vector }\label{a:pb}

The direct problem is given by Eq.~(\ref{eq-modele-direct}), with the slope and scintillation noise covariance matrix $C_d$ taken here equal to zero for simplicity.
Indeed $C_d$ appears only as a bias in the direct model, which is easily subtracted by the joint estimation with the $C_n^2$ profile as described in Sect.~\ref{coslidar}.
In our simulations hereafter, we compute the reduced data $C_{mes}$ with a known $C_n^2(z)$ profile and a given occurrence of convergence noise $u$. The known $C_n^2(z)$ profile, called the true profile, is incidentally taken as the profile reconstructed from the experimental data at 14:15 on 14 September 2012 shown on Fig.~14a of the paper.
Occurrences of convergence noise $u$ are simulated so as to respect the relevant covariance matrix $C_{conv}$. To this aim, the covariance matrix is diagonalized and a white noise is drawn in the $C_{conv}$ eigenvector basis.

\subsection{Generalized Cross Validation} \label{a:meth}

The Generalized Cross Validation method allows choosing in an unsupervised fashion a good value for the quadratic regularization hyper-parameter.
It is a classical method which can be applied to any linear ill posed problem~\cite{golub_generalized_1979}.
Generalized Cross Validation is so named because it approximates the average error that would be made by predicting one data sample using all the others, for the current value of the hyper-parameter $\mu$. It then chooses the value of~$\mu$ that minimizes this average prediction error.

\subsubsection{GCV Method}
The GCV method computes a function $V(\mu)$ for a range of values of the hyper-parameter $\mu$ and determines for which $\mu$ value, $V(\mu)$ is the lowest.
The GCV function $V(\mu)$ is generally given for a generic MAP criterion with a homogeneous white noise and a white quadratic regularization, \revision{also called Tikhonov regularization, } in the form:
 \begin{equation} \label{eq:forme_stand}
        J(x) = ||Ax-y||^2
        +\mu ||x||^2 \ , 
   \end{equation}
 where $y$ is the data and $x$ the unknown. 
The GCV function reads~\cite{golub_generalized_1979,bertero_introduction_1998,Demoment-l-08a}:
    \begin{equation}\label{GCV}
        V(\mu) =  \frac{||A\hat{x}(\mu)-y||^2}{[\mathrm{trace}(I-G(\mu))] ^2} \ ,
    \end{equation}
with $\hat{x}(\mu)=R(\mu)y$ the MAP solution for hyper-parameter $\mu$, $R(\mu)=(A^TA + \mu.I)^{-1}A^T$ the MAP reconstruction matrix, $I$ the identity matrix, and $G(\mu)= A R(\mu)$~\cite{golub_generalized_1979,Demoment-l-08a}. Note that by using the alternate form of the reconstruction matrix $R(\mu)=A^T(AA^T+\mu I)^{-1}$, the expression of $G(\mu)$  takes another, equivalent form: $G(\mu)=A A^T(AA^T+\mu I)^{-1}$~\cite{bertero_introduction_1998}.



\revision{The regularization of Eq.~\ref{eq:forme_stand} is the same as the one we use in Eq.~\ref{eq-Jprior-white-homog-quadratic}, but the data fidelity term is a simple least-squares and thus does not take into account the inhomegeneous and correlated noise at hand in our problem.}
By a change of variable, developed by~\cite{krakauer_using_2004}, and taking $x$ as the sought $C_n^2$ profile, we can reformulate our metric $J_{MAP}$ of Eq.~(\ref{eq:JMAP}), which accounts for an inhomegeneous and correlated noise,  so that it takes the form of~Eq.~(\ref{eq:forme_stand}). 
The change of variable is the following:
 \begin{equation}
        A=-C_{conv}^{-1/2}M \ ,
    \end{equation}
    
     \begin{equation}
        y=-C_{conv}^{-1/2} C_{mes} \ .
    \end{equation}
    
\begin{figure}[!htb]
    	\begin{center}
		\begin{tabular}{c}
		\includegraphics[width=1\textwidth]{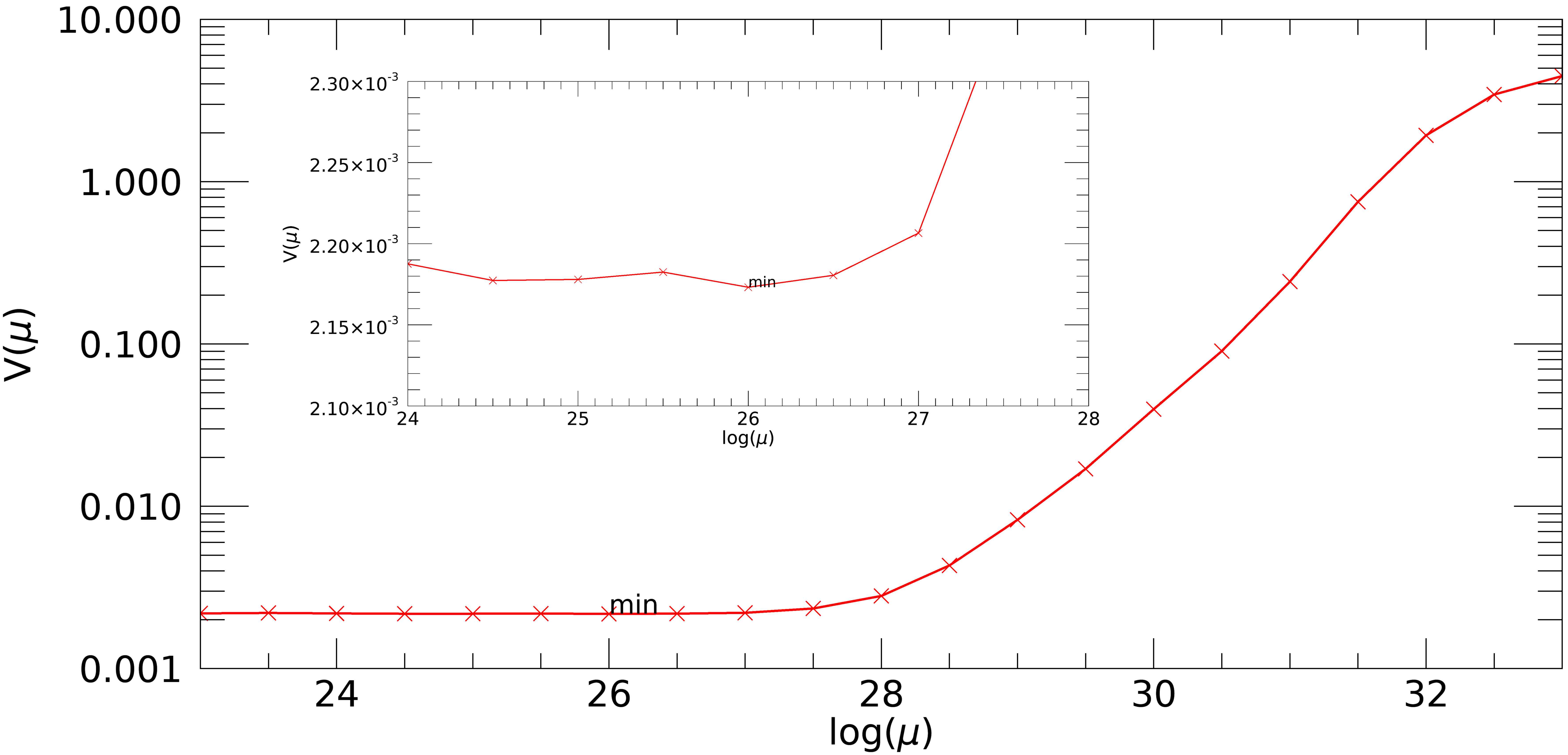}
		\end{tabular}
	\end{center}
    \caption{GCV function versus the regularization hyper-parameter $\log(\mu)$, with a focus on the minimum. We indicate the location of the minimum: $\log(\mu)$ = 26.}
    \label{figA:GCV}
\end{figure}

Figure \ref{figA:GCV} shows the GCV function $V(\mu)$ for a range of values of the hyper-parameter $\mu$, with $\log(\mu)$ ranging from 23 to 33 with a 0.5 step. The minimum of $V(\mu)$ is obtained for $\log(\mu) = 26$.
    
\subsubsection{Performance of the hyper-parameter selection}\label{A:hyp_selec}
  
To evaluate the quality of the hyper-parameter given by the GCV method, we evaluate the quadratic distance between the reconstructed profiles and the true profile, defined as the Mean Squared Error (MSE):
  	\begin{equation}
	  MSE(\mu) = \frac{\sqrt{\langle|C_n^2(z)- C^2_{n, true}(z)|^2\rangle_z}}{\langle C^2_{n,true}(z)\rangle_z} \ ,
	\end{equation}
	
where $\langle\cdot\rangle_z$ denotes averaging on $z$.

\begin{figure}[!htb]
    \begin{center}
	\includegraphics[width=1\textwidth]{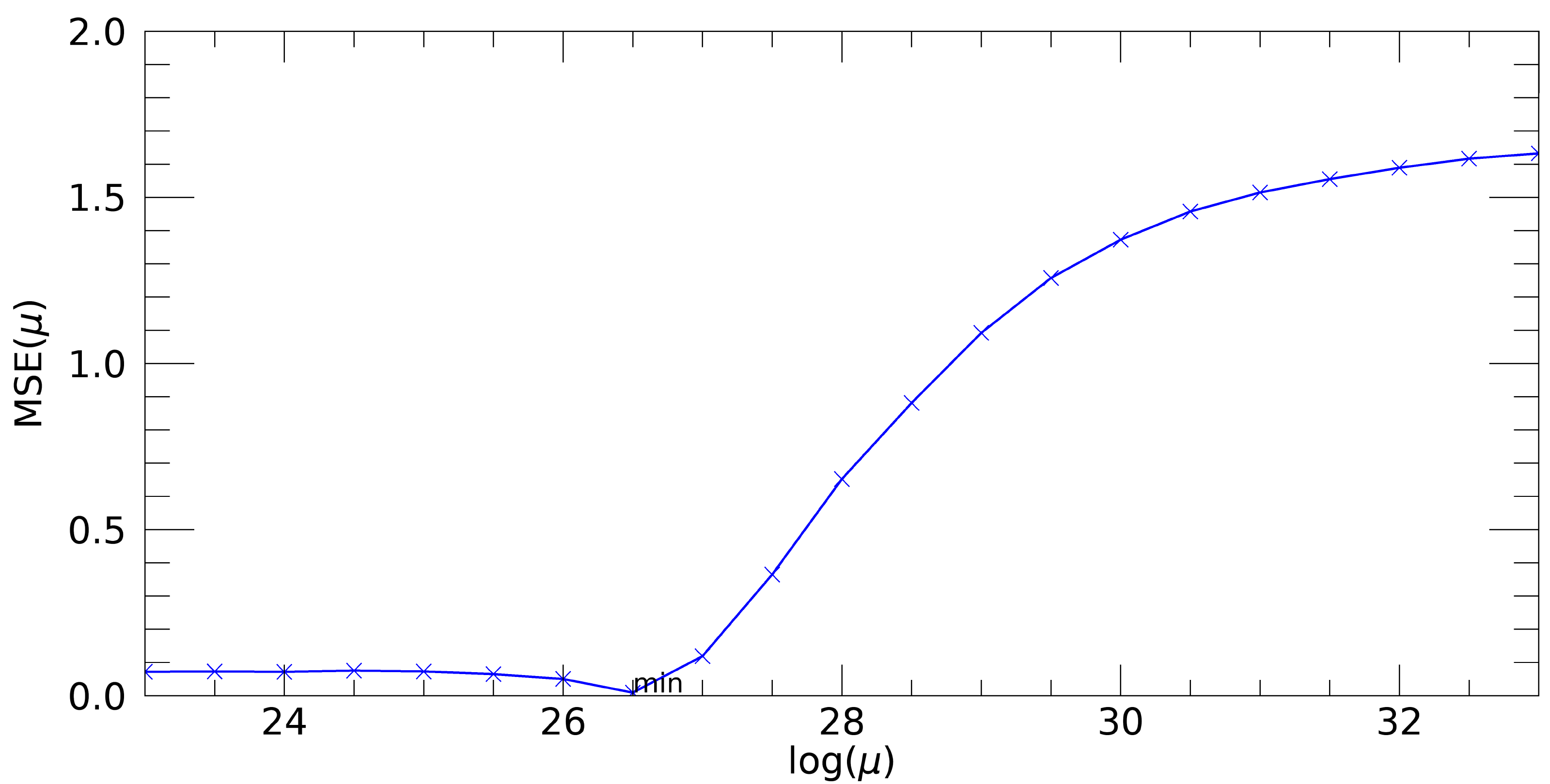}
	\end{center}
    \caption{MSE versus the logarithm of the regularization hyper-parameter $\mu$. The minimum is obtained for $\log(\mu)$ = 26.5.}
    \label{figA:MSE}
\end{figure}

Figure \ref{figA:MSE} shows the MSE of the reconstructed profile as a function of $\log(\mu)$. The minimum error is reached for the optimum value $\log(\mu)=26.5$---a ``best tuning'' value that is unreachable in practice with experimental data, because the true profile is unknown with experimental data. The value proposed by GCV ($\log(\mu)=26$) is thus very close to this optimum value without requiring the knowledge of the true profile.

\begin{figure}[!htb]
	\begin{center}
	\includegraphics[trim = 2.2cm 1.4cm 3.1cm 1.7cm, clip,width=1\textwidth]{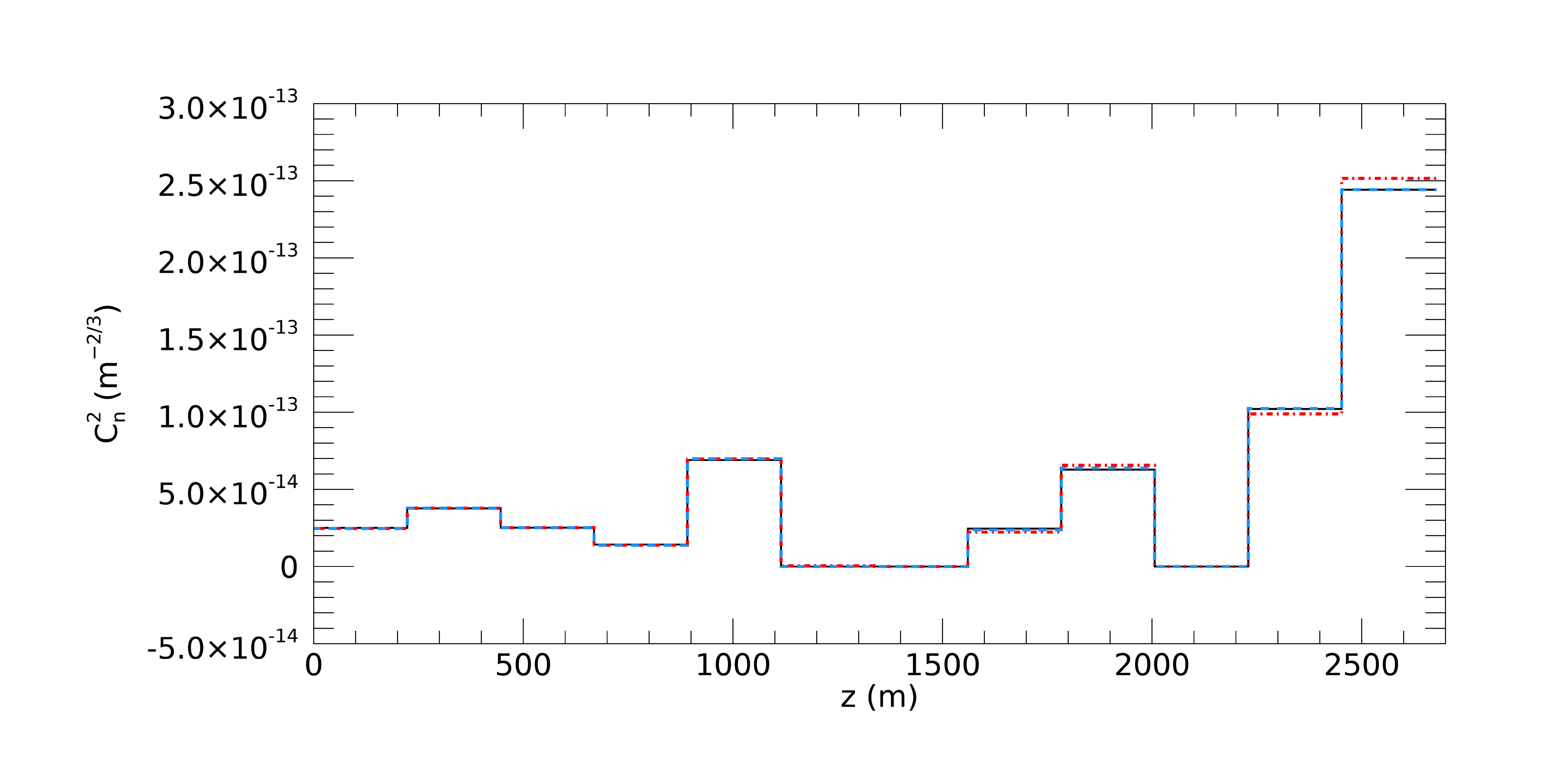}
	\end{center}
    \caption{$C_n^2$ profile reconstructions, with $\log(\mu) = 26$ in red dotted line and  $\log(\mu) = 26.5$ in blue dotted line. \revision{The true profile is drawn with a black line.}}
    \label{figA:P_diff_mu}
\end{figure}

Figure \ref{figA:P_diff_mu} shows the two reconstructed profiles for $\log(\mu)=26$ and $\log(\mu)=26.5$. We notice that they are both very close  to the true profile and that the only noticeable differences between the GCV-assisted reconstructed profile and the true one concern slices that are close to the sources. This is consistent with the results shown in Sect.~\ref{Cn2_profile}, where both the computed error bars and the sensitivity of the reconstruction to the value of $\mu$ are larger close to the sources.

In order to further validate the GCV method we simulate ten occurrences of the convergence noise to obtain ten occurrences of noisy reduced data. For each, we compute the GCV function and we reconstruct the $C_n^2(z)$ profile for the value of the hyper-parameter $\mu$ that minimizes the GCV function.
In practice during this campaign the $C_n^2$ has never been higher than $10^{-12.5}$, thus we take 25 like a lower bound of the $\log(\mu)$ range. 
The averaged squared difference between the GCV assisted $\log(\mu)$ and the ``best tuning''$\log(\mu)$ is equal to 0.76.

These results show the relevance of using the GCV method to provide a satisfactory hyper-parameter adjustment.

\subsection{Comparison of the MAP error bars and the empirical error bars}
\label{A:resul}

With the ten occurrence of convergence noise and associated reconstructed $C_n^2$ profiles of Sect.~\ref{A:hyp_selec} we can assess the quality of the error bars. 
We compare the average of the ten 1$\sigma$ MAP error bars -- computed with Eq.~(\ref{eq-error-covariance}) -- to the average of the empirical error bars ($EEB$), we define as:
\begin{equation}
    EEB(z) = \sqrt{\langle(C_n^2(z)- C^2_{n,true}(z))^2 \rangle_u} \ ,
\end{equation}
where $\langle . \rangle_u$ denotes the average on the ten reconstructions.

\begin{figure}[!htb]
	\begin{center}
	\includegraphics[trim = 2.2cm 1.8cm 3.1cm 1.7cm, clip,width=1\textwidth]{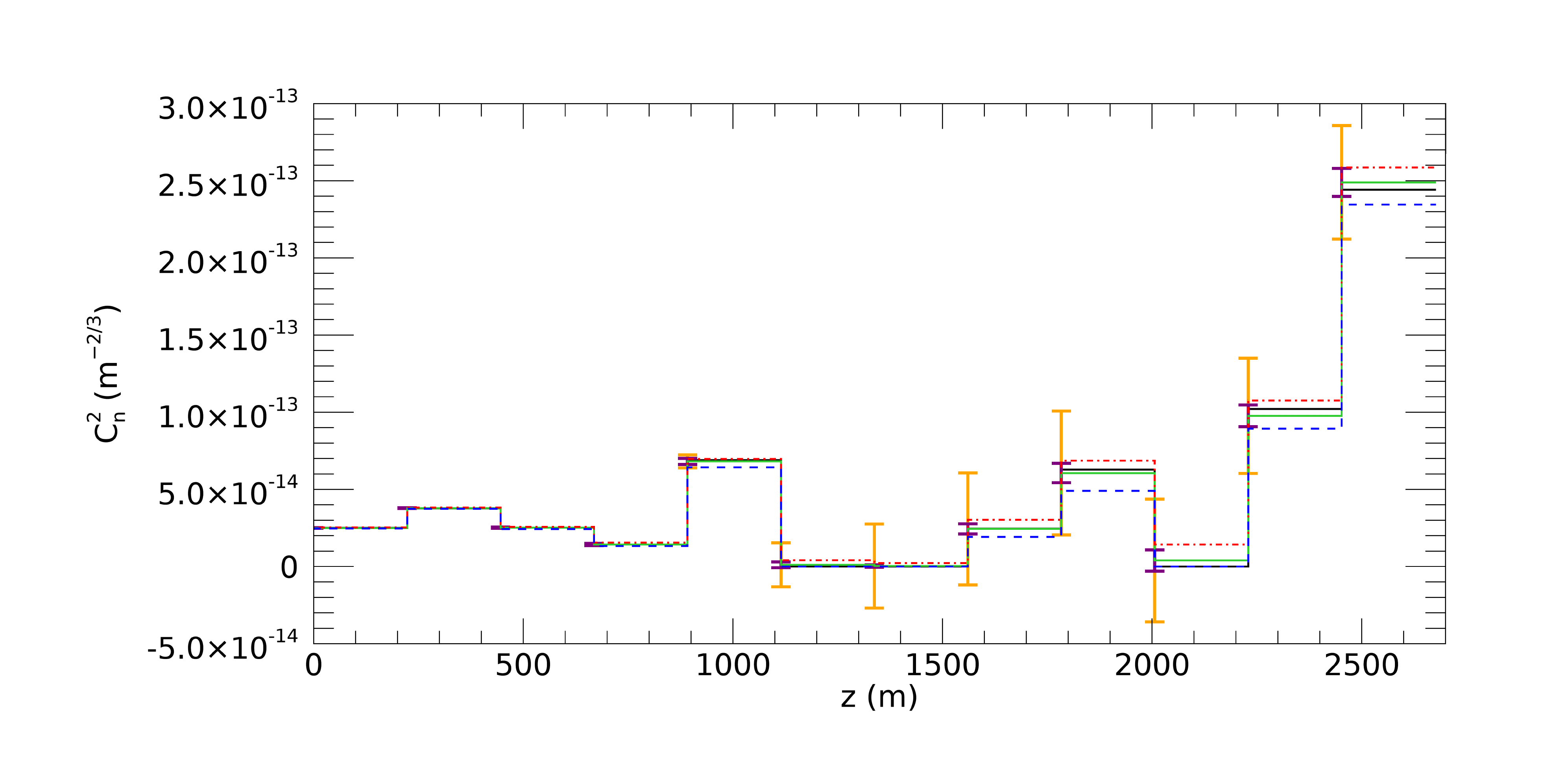}
	\end{center}
    \caption{Average of 10 reconstructed $C_n^2$ profiles (green line) with the minimum (blue line) and maximum (red line) among the 10 $C_n^2$ values for each slice. The error bars in yellow and purple represent respectively the average of the ten MAP error bars and the empirical error bars. The true profile is drawn in black line.}
    \label{figA:moy}
\end{figure}

Figure~\ref{figA:moy} shows the average of the ten $C_n^2(z)$ GCV-assisted unsupervised reconstructed  profiles (green line), along with the maximum (red line) and the minimum (blue line) values of the 10 reconstructed $C_n^2(z)$ profiles. The error bars in yellow represent the average MAP error bars, and the purple ones are the empirical error bars.
We first observe that the averaged $C_n^2(z)$ profile is very close to the true profile.
Additionally, we notice that the averaged MAP errors bars (in yellow) are larger than the empirical ones (in purple). 
Our interpretation is that the MAP error bars do not take into account the fact that a positivity constraint on the sought $C_n^2$ profile is applied during the reconstruction and acts as an additional regularization, which reduces the variability of the reconstructions.

\revision{Note that although the 1-sigma MAP error bars and the empirical error bars are represented as symmetrical tick marks in Fig.~\ref{figA:moy}, one should keep in mind that the histogram of the reconstruction error measured in the EEB is \emph{not} symmetrical for slices with $C_n^2(z)$ close to zero, due to the positivity constraint applied in the reconstructions\footnote{%
\revision{As suggested by the minimum and maximum values of the reconstructed profiles (in blue and red respectively in Fig.~\ref{figA:moy}), the histograms of the reconstructed $C_n^2$ values are quite symmetrical when the true $C_n^2(z)$ is far from zero, e.g. for the 12th and rightmost slice of Fig.~\ref{figA:moy}, and very asymmetrical when $C_n^2(z)$ is zero, e.g. for the 6th slice.}
}.}

The computed MAP error bars are thus conservative in practice and can therefore be applied to experimental data with confidence.

\subsection{Conclusion}
The GCV method gives satisfactory results for the unsupervised adjustment of the hyper-parameter, indeed it is close to the best hyper-parameter than can be reached.
The computed MAP error bars are shown to be conservative in practice.
This framework of GCV method and MAP error bars is applied to obtain the $C_n^2(z)$ profiles presented in Sects.~\ref{data-check}~and~\ref{Cn2_profile}.


\bibliography{Acronymes,EnglishAcronyms,Articles}

\end{document}